\documentstyle[11pt]{article}

\input{psfig}
\def\A{{\cal{A}}}

\def\boldlarrow {\mbox{\boldmath$\scriptstyle\leftarrow$\unboldmath}}
\def\boldrarrow {\mbox{\boldmath$\scriptstyle\rightarrow$\unboldmath}}
\def\boldrlarrow{\mbox{\boldmath$\scriptstyle\leftrightarrow$\unboldmath}}
\def\boldlsubarrow {\mbox{\boldmath$\scriptscriptstyle\leftarrow$\unboldmath}}

\def\bra#1{ \langle{#1}| }
\def\braket#1#2{ \langle{#1}|{#2}\rangle }
\def\opket#1#2 {  {#1} |{#2}\rangle }
\def\braopket#1#2#3{ \langle{#1}| {#2} |{#3}\rangle }

\def\chihat{\hat{\chi}}

\def\d{{\rm d}}
\def\D{{\cal D}}
\def\Deltahat{\hat{\Delta}}
\def\dim{d}

\def\dfdxdy#1#2#3{  {   {\partial^2 #1} \over
            {\partial\mathstrut{#2}\partial\mathstrut{#3}} } }
\def\dfdxdyh#1#2#3{  {   {\partial^2 #1} /
            {\partial\mathstrut{#2}\partial\mathstrut{#3}} } }
\def\dydxh#1#2{ \partial{#1} / \partial {#2} }
\def\dydxv#1#2{ {{\partial #1} \over {\partial #2}} }
\def\det{{\rm det}\,}

\def\E{{\cal E}}

\def\Gh{\hat{G}}

\def\Gtil{{\cal G}}

\def\ha{{1 \over 2}}
\def\H{{\cal H}}
\def\hh{\hat{h}}
\def\Hh{\hat{H}}
\def\Im{{\rm Im}}

\def\K2{{\cal K}}
\def\ket#1{ |{#1}\rangle }

\def\L{\stackrel{\!\!\boldrlarrow}{\D_n}}

\def\Lp{\stackrel{\!\!\!\!\boldrlarrow}{\D_{n'}}}
\def\Lpp{\stackrel{\!\!\!\!\!\!\boldrlarrow}{\D_{n''}}}
\def\Ls{{\!L}}

\def\myspace{\,}

\def\O{{\cal O}}
\def\off{(\Delta E/2)}
\def\Orient{R}
\def\p{{\bf p}}
\def\phat{\hat{p}}

\def\psil{\stackrel{\raisebox{-3pt}[0pt][0pt]{\boldlarrow}}{\psi}}
\def\psir{\stackrel{\raisebox{-3pt}[0pt][0pt]{\myspace\boldrarrow}}{\psi}}

\def\psisr{\stackrel{\raisebox{-3pt}[0pt][0pt]{\myspace\boldrarrow}}
		{\psi}^{\raisebox{-3pt}[0pt][0pt]{\mbox{$\scriptstyle*$}}}}
\def\q{{\bf q}}

\def\R{{\sigma}}

\def\Rs{{\!R}}
\def\s{{\bf s}}
\def\Sigl{\stackrel{\raisebox{-2pt}[0pt][0pt]{\boldlarrow}}{\Sigma} }
\def\Sigr{\stackrel{\raisebox{-2pt}[0pt][0pt]{\boldrarrow}}{\Sigma} }
\def\Siglsub{\stackrel{\raisebox{-1.5pt}[0pt][0pt]{\boldlsubarrow}}{\Sigma} }

\def\T{{\cal T}}

\def\Thhat{\hat{P}}

\def\Tim{{\cal F}}
\def\Tr{{\rm Tr}}
\def\Trl{F}
\def\tun{{\rm tun}}
\def\U{{\cal U}}

\def\vhat{\hat{v}}
\def\Vi{V_{\rm inner}}
\def\Vo{V_{\rm outer}}

\def\Wl{\stackrel{\raisebox{-2pt}[0pt][0pt]{\boldlarrow}}{\cal W}}

\def\Wt{{\cal W}}
\def\x{{\bf x}}
\def\spinor#1#2{\left( \begin{array}{c} {#1}\\
			                {#2} \end{array} \right)}

\def\zetab{\mbox{\boldmath$\zeta$\unboldmath}}
\def\zetabs{\mbox{\boldmath$\scriptstyle\zeta$\unboldmath}}

\def\argue{(\x,\x',E)}
\def\Ghd{\Gh^\dagger}

\def\Gt{\Gh_\tun}

\def\sbar{\big/}
\def\dbar{ {\,\sbar\!\sbar} }

\begin{document}
\baselineskip 15pt

\title{A Matrix Element for Chaotic Tunnelling Rates and Scarring Intensities}
\author{Stephen C. Creagh${}^{a,b}$ and Niall D. Whelan${}^{a,c}$}
\date{}
\maketitle

\vspace{-25pt}
\begin{center}
{\it

$^{a}$Division de Physique Th\'{e}orique$^*$,
Institut de Physique Nucl\'{e}aire 91406, Orsay CEDEX, France.
\newline
$^{b}$Service de Physique Th\'eorique,CEA/Saclay, 
91191, Gif-sur-Yvette CEDEX, France.
\newline
$^{c}$Department of Physics and Astronomy, Mcmaster University,
Hamilton, Ontario, Canada L8S~4M1.
}
\end{center}

\vspace*{10pt}

\begin{abstract}
It is shown that tunnelling splittings in ergodic double wells and
resonant widths in ergodic metastable wells can be approximated as
easily-calculated matrix elements involving the wavefunction in the
neighbourhood of a certain real orbit. This orbit is a continuation of
the complex orbit which crosses the barrier with minimum imaginary
action. The matrix element is computed by integrating across the orbit
in a surface of section representation, and uses only the wavefunction
in the allowed region and the stability properties of the orbit. When
the real orbit is periodic, the matrix element is a natural measure of
the degree of scarring of the wavefunction. This scarring measure is
canonically invariant and independent of the choice of surface of
section, within semiclassical error. The result can alternatively be
interpretated as the autocorrelation function of the state with
respect to a transfer operator which quantises a certain complex
surface of section mapping. The formula provides an efficient
numerical method to compute tunnelling rates while avoiding the need
for the exceedingly precise diagonalisation endemic to numerical
tunnelling calculations.
\end{abstract}

%\pacs{PACS numbers: 03.65.Sq, 05.45.+b}

\section{Introduction} 
\label{introduction}

\noindent Tunnelling in one dimension is an extremely well-studied and
understood problem \cite{bermount}. Using standard WKB methods,
together with appropriate uniform approximations, one can accurately
approximate such quantities as subbarrier transmission amplitudes,
energy level splittings in double wells and resonance lifetimes in
quasi-bound systems. In general this is not true in higher dimensions.
The problem is often enormously more difficult, for the simple reason
that the underlying classical mechanics is qualitatively richer.

This situation is especially evident in problems with classically
chaotic motion. In such cases, there is no simple constructive theory
for wavefunctions and any incorporation of tunnelling effects must
confront this fact. It is certainly impossible to write simple closed
expressions for a tunnelling rate associated with an individual state
along the lines of what exists in one dimension.  In this paper we
develop a formula which comes as close as possible within these
limitations. We concentrate on energy level splittings between states
in a symmetric double well, though the calculation for resonance
widths is also summarised. Our calculation is built upon an approach
developed by Auerbach and Kivelson \cite{AK} and by Wilkinson and
Hannay \cite{wh}, henceforth referred to as AK and WH respectively.
They use semiclassical Green's functions, constructed from complex
trajectories, to extend wavefunctions from allowed to forbidden
regions and thence to calculate spectral tunnelling effects.  We show
how this method can be used to express the splitting as a matrix
element which measures the concentration of the wavefunction in the
neighbourhood of a particular classical trajectory emerging from the
optimal tunnelling route.

The formula assumes that the wavefunction is known (possibly
following numerical diagonalisation), but otherwise uses quantities
that are straightforwardly calculated from classical dynamics.  The
technical benefit is that, while a completely numerical determination
of the splitting requires that diagonalisation be performed to a 
precision at least comparable with the energy level splitting (which can
be extreme), this matrix element requires only that the wavefunction
be calculated to a precision comparable with standard semiclassical
errors. Perhaps more importantly, there is a theoretical benefit in
that tunnelling rates are simply and directly related to the
the morphology of the wavefunction in the allowed region. This can 
be used as a starting point for statistical analyses and also for 
semiclassical analysis using Green's functions etc.

When the trajectory underlying the matrix element is periodic, the
splitting is nothing other than a scarring weight of the
wavefunction. Scarring \cite{hellerscars}, or the excessive
accumulation of certain eigenstates around periodic orbits, has become
a common means of characterising chaotic wavefunctions. Intuitively,
one expects exceptionally high tunnelling rates to occur in states
having a large amplitude along the optimal tunnelling route. The
matrix element quantifies this idea. It measures the amplitude of the
wavefunction along a certain real trajectory which connects to what we
will call the bounce orbit --- an imaginary-time complex trajectory
which crosses the barrier with minimal imaginary action. One often
finds that the bounce orbit lies on a symmetry axis, in which case its
real extension is periodic and the terminology of scarring is
appropriate. The formula works even if the orbit is not periodic,
however, though the scarring interpretation is then weaker.

There has recently been a lot of interest in the current-voltage
characteristics of tunnelling diodes in the presence of crossed
electric and magnetic fields \cite{experiment1,experiment2}. In those
systems it has been argued that the tunnelling current is strongly
mediated by scarred states. In this paper we quantify that very
connection between scarring and tunnelling but in contexts where the
tunnelling segment plays an active role in dynamics (up to now,
tunnelling diodes have been treated assuming the barriers to be
sufficiently thin that tunnelling can be incorporated simply using
reflection and transmission coefficients). There are some formal
similarities between our results and certain calculations for
tunnelling diodes --- our configuration space formulation to that of
\cite{RB} and the phase space formulation to that of \cite{SN}. There
are also differences however. We assume the wave function is highly
excited and fluctuating strongly and get a matrix element with a
strongly localised Green's function. In \cite{RB,SN} one finds the
contrary: the wave function is localised because incoming electrons
are assumed to be in an approximate ground state and the Green's
function, associated with real dynamics, is strongly oscillatory.

The paper is divided as follows. An important element in getting
simple and universal expressions for the splitting is to normalise
it by an average background associated with the bounce orbit. In
\cite{us} we derived an expression for the average splitting.
Controlled by the imaginary action of the bounce orbit, this average
varies strongly with energy and with system parameters. Dividing each
splitting by the average splitting at that energy leaves a set of
rescaled splittings which can usefully be compared between
states lying in very different parts of the spectrum. We begin in
section~\ref{theoryforthemean} by reviewing the most important
elements of this calculation.

In section~\ref{ahybridapproximation} we review the calculations of AK
and WH and give a derivation of the main matrix element.  The review
is necessary because certain technical aspects of the calculation,
having to do with choice of Green's functions {\it etc.}, must be
stressed in order to understand our result. In addition we implement
the formula somewhat differently; we consider the value of the
wavefunction along a cut inside the well rather than on the energy
contour as done in WH or along a cut in the forbidden region as in
AK. Having restated the result, an interpretation is developed as a
matrix element in a Poincar\'e-section representation similar to that
used in the transfer operator method of Bogomolny
\cite{bogomolnytrans}. In a Wigner-Weyl calculus restricted to the
Poincar\'e section, the rescaled splitting is simply the overlap of
the Wigner function of the state with a Gaussian centered on the real
extension of the bounce orbit.  The exponent of this Gaussian is
directly related to the complex stability matrix along this tunnelling
orbit and is a quadratic function on the Poincar\'e section.
In section~\ref{resonancescase} we outline how these results may be
translated to the calculation of resonance widths in metastable wells.
In section~\ref{Transop} the general results are given a formal 
interpretation in terms of transfer operators and this is then used
as the basis for a  discussion of the conditions under which positivity 
of the splittings is expected.

In section~\ref{thesystem} we use the theory to calculate splittings of 
a model
system and compare to the exactly determined values. The agreement is
impressive; for the bulk of the splittings, the theory is correct to
within a few percent.

\section{Extracting the Mean} 
\label{theoryforthemean}

\begin{figure}[h]
\vspace*{0.5cm}\hskip 2.5cm 
\psfig{figure=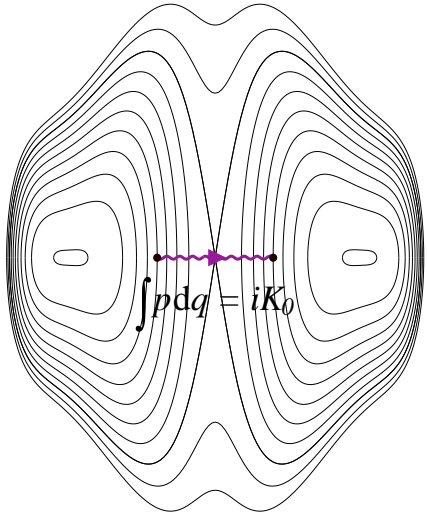,height=2.5in}
\vspace*{0.5cm}
\caption
{\small Here are potential contours of a double well with
chaotic dynamics. This potential, corresponding to Eq.~(\ref{fop})
with $\mu=0.1$, will be used in numerical investigation of the
results. Also represented is the complex bounce orbit crossing 
the barrier in imaginary time $-i\tau_0$ and defining the  tunnelling 
action $K_0$. Linearisation of dynamics about this orbit defines a 
complex monodromy matrix $W_0$. The classical data $K_0$ and $W_0$ 
together determine the mean splitting function $f_0(E)$.}
\label{potview}
\end{figure}

\noindent We consider double wells with predominantly chaotic
dynamics. Our detailed calculations are for potentials $V(x,y)$ in two
dimensions, though the general results and conclusions apply to higher
dimensions. A specific example of such a system is provided by the 
potential,
\begin{equation} \label{fop}
          V(x,y) = (x^2-1)^4 + x^2y^2 + \mu y^2,
\end{equation}
which will be used in section~\ref{thesystem} for explicit numerical
testing of our results. For positive $\mu$, this potential forms 
two wells, symmetric in $x$ and separated by a saddle at the origin 
with energy $E=1$, as illustrated in Fig.~\ref{potview}.

In a one-dimensional double well, the energy level splittings $\Delta
E_n$ vary monotonically from one doublet to the next (doublets are
indexed by $n$). In fact, the splitting is a simple, smooth function
of the mean energy $E_n$ of the doublet. In a chaotic multidimensional
well, this picture suffers a qualitative change --- while in any given
energy range there is a typical scale for splittings, individual
splittings vary quasi-randomly from one doublet to the next.  The
primary purpose of this paper is to discuss these fluctuations in the
tunnelling rate and to relate them to the properties of individual
wavefunctions. In order to do this in a coherent way, it is useful
first to calculate the typical size of splittings in any given energy
interval and to use this to rescale the calculation so that
fluctuations are always of the order of unity. That is the object of
this section.  We define a simple function of energy which predicts
the mean value of the splittings. This is then used to define a
rescaling of the splittings whose fluctuations are essentially
universal.

In \cite{us} we introduced the splitting-weighted density of states
\begin{equation}\label{deff}
               f(E) = \sum_n \Delta E_n \delta(E-E_n).
\end{equation}
Knowledge of this spectral function gives a complete specification of
the tunnelling properties of the system. A semiclassical expansion of
$f(E)$ was found in terms of complex periodic orbits crossing the
barrier.  Here we limit ourselves to the dominant term,
\begin{equation}\label{trunc}
            f_0(E)  =  {1\over\pi}
                       {e^{-K_0/\hbar}\over \sqrt{
			(-1)^{d-1}\chi_\R \mbox{det} (W_0-I)}},
\end{equation}
which furnishes the average properties of the splittings.
It is calculated from what is commonly called the ``instanton'' or 
``bounce'' orbit, illustrated in Fig.~\ref{potview}. This is a complex 
orbit which crosses the barrier  with imaginary
momentum and real position in an imaginary time $t=-i\tau_0$ 
and with an imaginary action $S=iK_0$; it can be understood in terms 
of a real orbit of
the system in which the potential is turned upside down.  For
two-dimensional potentials which are symmetric in $y$, the orbit
simply runs along the $x$-axis bouncing between the two energy
contours. $W_0$ is then a complex $2\times 2$ monodromy matrix
linearising motion transverse to the orbit, which can be put in the
form
\begin{equation}\label{defW0}
	     W_0 = \left( \matrix { \phantom{i}w_0 &           -iu_0 \cr 
		                              iv_0 & \phantom{-i}w_0  }  
							\right)    .
\end{equation}
The real numbers $(w_0,u_0,v_0,w_0)$ are simply the matrix elements of the
real monodromy matrix in the inverted problem. Equality of the diagonal 
elements follows from a symmetry of the orbit under time-reversal 
(this structure is altered somewhat in the presence of a magnetic field 
and will be discussed more fully in section~\ref{SCGsec}). In any 
case, even though $W_0$ is complex, the amplitude term $\det(W_0-I)$ 
is real, as are $W_0$'s eigenvalues. Finally, $d$ is the dimension of 
the system and $\chi_\R$ is simply a sign fixed by the nature of 
the symmetry operation $\R$ that underlies the double well. 
When $\sigma$ is an orientation reversing transformation of configuration
space, $\chi_\R=+1$ and in the orientation preserving case, $\chi_\R=-1$.
In particular, $\chi_\R=+1$ in the case where $\sigma$ is 
a reflection in one coordinate, as in the problem used to 
illustrate the calculation here. In practice, it
can be determined simply by demanding that the argument of the square 
root be a positive number.

The function $f_0(E)$ yields the average splittings in the sense that
if $f(E)$ is averaged within a sufficiently large window, the
contributions of other complex periodic orbits to $f(E)$ are
suppressed. This means that the average value of splittings,
when measured in units of the mean 
spacing between doublets, is 
$\langle\rho_0(E_n)\Delta E_n\rangle = f_0(E)$.
We denote by $\rho_0(E)$ the average or Thomas-Fermi density of 
states within one well (computed for a given parity in $y$ if 
appropriate) --- its inverse is the mean spacing between doublets. 
Explicit numerical
verification that $f_0(E)$ describes the mean behaviour can be found in
\cite{us}.

Note that $f_0(E)$ typically increases rapidly with energy and
therefore so do the typical values of the splittings. It is more
convenient to work instead with normalised splittings $y_n$, defined
by dividing by the local average,
\begin{equation}\label{defy}
		y_n  =  {\rho_0(E_n)\over{f_0(E_n)}} \Delta E_n.
\end{equation}
These have average value of unity, by construction, and can usefully 
be compared between different parts of the spectrum, or even between
different  systems.

\section{Derivation of the Matrix Element for Splittings} 
\label{ahybridapproximation}

\noindent In investigating fluctuations in the tunneling rate, 
one option is to include, in the formalism alluded to in the previous
section, complex orbits with real parts to their actions.  Analysis of
this kind was initiated in \cite{us} and a systematic enumeration of
the most important such orbits will appear in \cite{us'}.  This
approach is most powerful on large energy scales, where relatively few
complex orbits suffice to reproduce extremely detailed collective
behaviour of tunnelling rates. When the tunnelling rate associated
with an {\it individual} level is required, however, practical
implementation of this program can be daunting due to the large number
of long orbits which must be calculated to reach the energy scale of a
typical level spacing. For this reason it is of value to pursue an
alternative approach, as we do in this section, which steers a course
between purely semiclassical and fully quantum calculation to specify
individual tunnelling rates.

Our discussion follows the methodology of AK \cite{AK} and WH
\cite{wh}, whereby each splitting is expressed as an overlap between
approximate states localised in each of the wells. As discussed in WH,
for chaotic systems, these local approximations are not found using
semiclassical approximations.  Instead, it is assumed that the
wavefunctions in the wells are found through some numerical
diagonalisation or possibly through some random matrix theory
modelling if we are interested in statistical properties of the
splittings. Once the wavefunction has been constructed in the wells,
the splitting is computed by extending the wavefunction under the
barrier using semiclassical approximations to the Green's function.

We give in this section an explicit derivation of the main formula.
The purpose of this is two-fold. First, it is important for practical
implementation to emphasise certain boundary conditions placed on the
complex classical trajectories used in the formalism, having to do
with the path that time takes in the complex plane. In addition, we
would like to emphasise certain coordinate-invariant properties of the
calculation which might be useful for application to more general
problems.  While broadly similar to the AK and WH results, the final
form of our calculation differs from theirs in certain technical
aspects, mostly having to do with exercising a freedom as to which
parts of the wavefunction are used.

\subsection{An abstract Herring's formula}

\noindent
The starting point is Herring's formula \cite{herring}.  In our
derivation we assume $\ket{L}$ and $\ket{R}$ to be symmetric and
antisymmetric combinations of odd and even eigenstates $\ket{\pm}$
which are respectively localised in the left and right wells. (While
deriving the the splitting for an individual doublet, we will drop the
doublet label $n$). We denote the symmetry operation which relates the
two wells (typically a reflection or an inversion) by the symbol
$\sigma$ and call $U(\sigma)$ the corresponding unitary operator.  We
then have $\ket{R}=U(\sigma)\ket{L}$. If we write the coupled
Schr\"odinger equation as
\begin{eqnarray}\label{defLR}
	\Hh\ket{L} & = & E\ket{L} - \off\,\ket{R}  \nonumber\\
	\Hh\ket{R} & = & E\ket{R} - \off\,\ket{L} ,
\end{eqnarray}
then the eigenenergies are $E^\pm = E\mp\Delta E/2$ and the splitting
$\Delta E = E^- - E^+$ is positive when the even state is at a lower
energy than the odd state.  Now let $\Thhat$ be any Hermitean operator
such that $\Thhat\ket{L}\approx0$ and $\Thhat\ket{R}\approx\ket{R}$.
A typical choice would be to let $\Thhat$ represent multiplication of
the wavefunction by the characteristic function $\chi_\Sigma(\x)$ of
the region to the right of some section $\Sigma$ (of codimension $1$)
which separates one well from the other. The precise form will not
play an important role here. Applying $\Thhat$ to each of the
equations above and taking matrix elements, we find
\begin{eqnarray}\label{halfHerring}
  \braopket{R}{\Thhat\Hh}{L} 
       	    & = & 
		E\braopket{R}{\Thhat}{L} 
	 	- \off\,\braopket{R}{\Thhat}{R}  \nonumber\\
  \braopket{L}{\Thhat\Hh}{R} 
       	    & = & 
		E\braopket{L}{\Thhat}{R} 
		- \off\,\braopket{L}{\Thhat}{L}.
\end{eqnarray}
Subtracting the complex conjugate of the first from the second we get

\begin{equation}\label{exactHerring}
	\braopket{L}{\left[\Thhat,\Hh\right]}{R} 
		=  \frac{\Delta E}{2}
	\left(\braopket{R}{\Thhat}{R}-\braopket{L}{\Thhat}{L}\right).
\end{equation}
The calculation is exact up to this point. We now take advantage of   
the approximations $\braopket{L}{\Thhat}{L}\approx 0$ and 
$\braopket{R}{\Thhat}{R}\approx 1$ to obtain,
\begin{equation}\label{genHerring}
	\Delta E \approx 2\,\braopket{L}{\left[\Thhat,H\right]}{R}.
\end{equation}
The error in this approximation is governed by leakage of the
wavefunctions of $\ket{L}$ and $\ket{R}$ across $\Sigma$.  This effect
is exponentially small so the above approximation is very accurate;
its error is almost certainly dwarfed by that of the next step, which
is to substitute approximations for $\ket{L}$ and $\ket{R}$, to be
discussed in the next subsection. Also, we will henceforth dispense
with using approximation signs for the splittings and similar
quantities, it being understood that any expression using
semiclassical arguments is an approximation.

This explicit expression for $\Delta E$, which is independent of
representation and holds for any form of the Hamiltonian, is a
generalised version of Herring's formula. By choosing $\Thhat$
suitably, it might even be used for calculating dynamical tunnelling 
splittings, where the division between the localised states is 
in phase space and not in configuration space. The important point 
is that it relates the splitting to an overlap of the localised 
states that is restricted to the forbidden region of phase space, 
where the Weyl symbol of $\Thhat$ is nonconstant. The form used in 
\cite{wh},
\begin{equation}\label{Herring}
	\Delta E \;=\; \hbar^2 \int_\Sigma \d s \; \left(
	 \psi_L^* \,\dydxv{\psi_R}{n}-\dydxv{\psi_L^*}{n}\,\psi_R\right)
\;\; ,
\end{equation}
is a special case obtained when $\Hh = \phat^2/2 +V(q)$ and 
$\Thhat$ is chosen to represent multiplication by the characteristic function
$\chi_\Sigma(\x)$ discussed above [it is a straightforward consequence of 
the relation $\left[\chi(q),\phat^2\right]=i\hbar (\chi'(q)\phat 
+ \phat \chi'(q))$ which holds for any function $\chi(q)$]. Here 
$\d s$ is a surface element along $\Sigma$ 
and the normal derivative  $\dydxh{}{n}$ is directed to the side of
$\Sigma$ containing $\ket{R}$. 
In this case we get an integral along
a lower-dimensional surface because $\chi_\Sigma(\x)$ rises
from $0$ to $1$ as a discontinuous step function. Had we used 
instead a smoothly rising function, the overlap would have 
appeared as a fully $d$-dimensional integral, where $d$ is the 
dimension of the system, restricted to the region around $\Sigma$
where $\nabla \chi_\Sigma(\x) \neq 0$. Other possibilities
might be possible if $\Thhat$ mixes momentum and position 
operators.

Matrix elements of the kind shown in Eqs.~(\ref{genHerring}) and
(\ref{Herring}) will repeatedly appear during the course of the
calculation, so it is useful at this point to introduce explicit
notation for them and point out those aspects which are important.  We
assume in general that we have a related pair $(\Thhat,\Sigma)$, where
$\Sigma$ is a section and $\Thhat$ is a Hermitean operator whose Weyl
symbol rises from $0$ to $1$ on passing from one side of $\Sigma$ to
the other.  In Herring's formula $\Sigma$ is in the forbidden region,
but in later incarnations, we will have conventional sections in the
allowed part of phase space. Defining the anti-Hermitean operator
\begin{equation}\label{defdelta}
	\Deltahat_\Sigma = \left[\Thhat,\Hh\right],
\end{equation}
we construct a sectional matrix element between two states $\ket{\phi}$ and 
$\ket{\psi}$ defined as,
\begin{equation} \label{defsecoverlap}
		\langle\phi\,\dbar_\Sigma\,\psi\rangle
			\equiv
		\braopket{\phi}{\Delta_\Sigma}{\psi}.  
\end{equation}
Note that $\langle\phi\,\dbar_\Sigma\,\psi\rangle^*
		=    -\langle\psi\,\dbar_\Sigma\,\phi\rangle$.
This change in sign might usefully be interpreted as a change 
in orientation of $\Sigma$, or the replacement of $\Thhat$ by $I-\Thhat$.
The sectional overlap defined in Eq.~(\ref{defsecoverlap})
will later in the calculation produce
matrix elements in the Poincar\'e section representation
introduced in \cite{bogomolnytrans} to define the transfer operator.
It is also similar to notation used in \cite{AK}. 
When $\Hh$ is 
of kinetic-plus-potential type and $\Thhat$ corresponds to a 
characteristic function $\chi_\Sigma(\x)$ as above, the matrix 
element becomes the standard surface integral
\begin{equation} 
		\langle\phi\,\dbar_\Sigma\,\psi\rangle
			= \frac{\hbar^2}{2} \int_\Sigma \d s \;
	 \left(\phi^* \,\dydxv{\psi}{n}-\dydxv{\phi^*}{n}\,\psi\right)
\;\; ,
\end{equation}
familiar from Green's identity, the unit normal being directed
to the side of $\Sigma$ where $\chi_\Sigma$ is $1$. 
Herring's formula,
\begin{equation}\label{secHerring}
        \Delta E = 2 \, \langle L\dbar_\Sigma\,R\rangle
\end{equation}
now gives the splitting as a sectional overlap of the left and right 
localised states, albeit in a region of phase space where dynamics is 
complex and the wavefunctions exponentially small.

\subsection{An abstract Green's identity}

\noindent
Herring's formula is formally interesting but not of direct use
without further analysis since explicit calculation of the
wavefunction in the forbidden region can be as difficult as finding
the splitting itself.  Following \cite{wh}, we simply assume the
wavefunction to be known in the allowed region and then use Green's
theorem to continue it into the forbidden region, using a
semiclassical approximation for the Green's function constructed
around complex classical trajectories.

\begin{figure}[h]
\vspace*{0.5cm}\hskip 2.5cm 
\psfig{figure=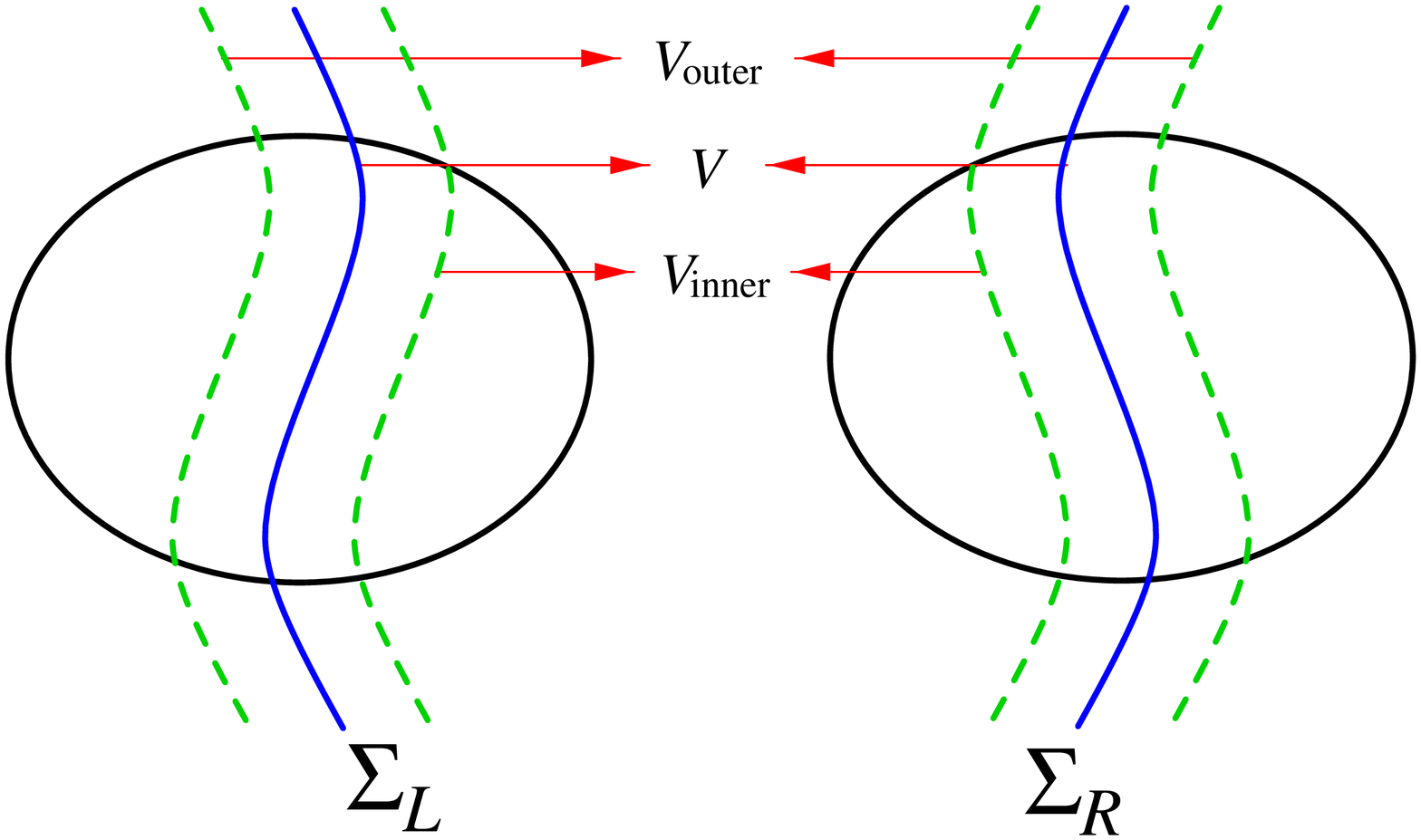,height=2.5in}
\vspace*{0.5cm}
\caption{\small A schematic representation of the three domains $\Vo$, 
$V$ and $\Vi$.
}
\label{Viofig}
\end{figure}

Let us define sections $\Sigma_L$ and $\Sigma_R=\sigma\Sigma_L$
cutting through the left and right wells respectively, as illustrated
in Fig.~\ref{Viofig}.  Let $V$ be the region between them. Let $\Vo$
be a larger region enclosing $V$ and $G\argue$ a Green's function
which satisfies Schr\"odinger's equation (with delta-function source)
as long as $\x$ and $\x'$ are within $\Vo$. We will not impose on
$G\argue$ that it satisfy the Schr\"odinger equation, with boundary
conditions, outside of $\Vo$. It may diverge at infinity, for
example. In fact, it will be important later that it not be the global
Green's function because it is then well-defined at quantised values
of $E$, which is where we will want to use it. This is analogous to
the use of free space Green's functions in the boundary-integral
method in billiards \cite{bim}. In our case, we will use a
semiclassical approximation for $G\argue$ in which trajectories are
discarded after they leave $\Vo$. The resulting function will then
satisfy the necessary equations inside $\Vo$ and has the added
advantage that, with a judicious choice of $\Vo$, only a finite number
of trajectories contribute for a given argument. We will further
suppose that we only ever want to reconstruct the wavefunction within
some region $\Vi$ enclosed by $V$. Let $\chi(\x)$ be a function which
is zero outside $\Vo$ and rises to unity passing through $\Sigma_L$ or
$\Sigma_R$ and is constant by the time $\Vi$ is reached.  The geometry
is sketched in Fig.~\ref{Viofig}.

That $G\argue$ is a Green's function inside $\Vo$ amounts to the
following in abstract notation:
\begin{equation}\label{localGreen}
  \chihat \Gh(E-\Hh) \chihat = \chihat (E-\Hh) \Gh \chihat = \chihat^2,
\end{equation}
where $\Gh$ and $\chihat$ are the abstract operators corresponding to
$G\argue$ and $\chi(\x)$. Note that $\chihat$ does not have an
inverse.  Let $\ket{\psi}$ be any state whose wavefunction satisfies
the Schr\"odinger equation in $\Vo$ with energy $E$, so that
$\chihat(E-\Hh)\ket{\psi}=0$. Then we can write the following two
relationships,
\begin{eqnarray}\label{preGreen}
    \chihat\Gh(E-\Hh)\chihat\ket{\psi}  &=&  \chihat^2\ket{\psi}\nonumber\\  
    \chihat\Gh\chihat(E-\Hh)\ket{\psi}  &=&  0.
\end{eqnarray}
Taking the difference gives,
\begin{equation}\label{longGreen}
       \chihat\Gh\left[\chihat,\Hh\right]\ket{\psi} = \chihat^2\ket{\psi}.
\end{equation}
Define the anti-Hermitean operator $\Deltahat=\left[\chihat,\Hh\right]$ 
in analogy with Eq.~(\ref{defdelta}). Then, provided $\x$ is in $\Vi$,
\begin{equation}\label{absGreen}
    \psi(\x) % =  \braopket{\x}{\chihat^2}{\psi} 
	       =  \braopket{\x}{\Gh\Deltahat}{\psi} 
	       =  \langle{\x}|G\!\dbar\psi\rangle.
\end{equation}
The overlap here is as in Eq.~(\ref{defsecoverlap}) except that the 
integration is along two sections $\Sigma_L$ and $\Sigma_R$ instead 
of the single section $\Sigma$. It is an abstract form of Green's 
identity capable of relating the wavefunction in the centre of the 
barrier to its values in the neighbourhoods of $\Sigma_L$ and $\Sigma_R$. 
Note that it was not fundamentally important in deriving this relation
that $\chihat$ represent multiplication by a function of position
--- it is only necessary that Eq.~(\ref{localGreen}) hold and 
that $\ket{\psi}$ locally obey the Schr\"odinger equation. This 
observation is helpful in generalising the result to the case of 
dynamical tunnelling where we would want to deduce the behaviour 
of the state in one part of {\it phase space} from its behaviour 
in  another. However, we will continue with the language of 
tunnelling between wells to avoid confusion of notation.

Let us now consider using this method to extend the wavefunction
$\psi_L(\x)$ from a neighbourhood of $\Sigma_L$ to the centre of the
barrier. We make two approximations, based on the premise that
tunnelling effects are required only to leading order in $\Delta E$.
First, Eq.~(\ref{absGreen}) is employed even though $\psi_L(\x)$ is
not, strictly speaking, an eigenfunction. To justify this we note that
inclusion of the off-diagonal term in Eq.~(\ref{defLR}) results in the
following correction to Eq.~(\ref{longGreen}),
\begin{equation}\label{longGreen2}
       \chihat\Gh\left[\chihat,\Hh\right]\ket{L} = \chihat^2\ket{L}
			- \frac{\Delta E}{2}\chihat\Gh\chihat\ket{R}.
\end{equation}
We claim that the additional term on the right hand side can be
neglected when evaluating the wavefunction near the centre of the
barrier.  This will be seen self-consistently, first by ignoring the
term and, following semiclassical approximation of the Green's
function later, noting that the imaginary action underlying
$\chihat\Gh\chihat\ket{R}$ is similar to that underlying
$\chihat^2\ket{L}$. Therefore, since $\Delta E$ is small compared to
other energy scales, the second term can be ignored.  The second
approximation is that, since $\psi_L(\x)$ is small in the right well,
the contribution to $\psi_L(\x)$ in Eq.~(\ref{absGreen}) from the
neighbourhood of $\Sigma_R$ is ignored.  The validity of each of these
assumptions can be explicitly verified in the case of WKB
approximations in one dimension. Thus we can recast
Eq.~(\ref{absGreen}) as
\begin{equation}\label{LRGreen}
    \psi_L(\x) = \braket{\x}{\Gh\dbar_{\!\Sigma_L}L}
\end{equation}
where $\dbar_{\Sigma_L}$ denotes a sectional overlap confined to 
$\Sigma_L$. With the obvious analog for $\psi_R(\x)$, we are now 
ready to invoke Herring's formula.

Denoting,
\begin{equation}\label{concatGdG}
	\Gt = \Ghd\left[\Thhat,\Hh\right]\Gh 
	    = \Ghd \!\dbar_\Sigma\, \Gh,
\end{equation}
which is a concatenation of advanced and retarded Green's functions 
along the section $\Sigma$, we have,
\begin{equation}
	\Delta E  =  -2\,
	\langle L \dbar_{\Sigma_L} \Gt \dbar_{\Sigma_R} R\rangle
\label{RLHerring}\label{rlherring}.
\end{equation}
The minus sign arises from the conjugation of $\psi_L(\x)$ following
calculation from Eq.~(\ref{LRGreen}) (cf. comment just below
Eq.~(\ref{defsecoverlap})).  Having substituted our continued
wavefunctions into Herring's formula, we get an expression for $\Delta
E$ involving three sectional integrals. There is one integral through
the centre from Herring's formula itself, and two more around the
surfaces of section coming from the wavefunction extensions. In our
notation the central integral becomes hidden in the definition of
$\Gt$. This is deliberate because, as will be seen in the next
section, this integral can be performed analytically within
semiclassical approximation. The matrix element in
Eq.~(\ref{RLHerring}) subsequently involves just two integrations,
along $\Sigma_L$ and $\Sigma_R$.

Formally, Eq.~(\ref{RLHerring}) is essentially the same as Eq.~(3.8)
of \cite{AK}. However, the formula is implemented differently in our
case --- Auerbach and Kivelson choose sections inside the forbidden
region whereas we choose them running through the middle of the
wells. In addition, the derivation of Eq.~(\ref{RLHerring}) could
equally well be valid for dynamical tunnelling if the operator
$\Thhat$ is chosen appropriately. The main complication of dynamical
tunnelling will come later, when the complex trajectories used to
construct $\Gt$ need to be calculated, in which case the structures
involved are less simple  than in the case of tunnelling
between wells.

In section~\ref{Transop} we will develop an interpretation for 
$\Gt$ as the transfer operator for a complex Poincar\'e mapping.
For conventional real Poincar\'e mappings, the transfer operator
is a finite-dimensional unitary matrix, quantising the classical 
first-return map \cite{bogomolnytrans}. We will define a complex 
mapping from $\Sigma_R$ to itself for which $\Gt$ acts like a 
quantisation in the same way (though it is not unitary in the complex 
case). Using the symmetry to identify the left and right 
wavefunctions, Eq.~(\ref{RLHerring}) is then formally like an 
autocorrelation function.

For the sake of concreteness, we write down the explicit forms taken by
these integrals when the sectional overlaps are of the form given in 
Eq.~(\ref{defsecoverlap}). First denote $\L = 
\stackrel{\!\raisebox{-2pt}[0pt][0pt]{\boldrarrow}}{\,\dydxh{}{n}}-
\stackrel{\!\raisebox{-2pt}[0pt][0pt]{\boldlarrow}}{\,\dydxh{}{n}}$,
the arrows indicating whether
the derivative acts on the function to the left or to the right. 
Then the central integral, defining 
$G_\tun\argue=\braopket{\x}{\Gt}{\x'}$, is
\begin{equation} \label{GRLint}
G_\tun\argue = \frac{\hbar^2}{2}
\int_\Sigma \d s'' \;\; G^\dagger(\x,\x'',E)\; \Lpp G(\x'',\x',E)
\end{equation}
and the outer integrals are
\begin{equation}\label{realHerring}
	\Delta E = -\frac{\hbar^4}{2} \int_{\Sigma_L} \d s 
			   	      \int_{\Sigma_R} \d s' \;\;
      \psi_L^*(\x) \,\L\, G_\tun\argue\,\Lp\, \psi_R(\x'). 
\end{equation}
Remember that our convention is that the normals to $\Sigma_L$ 
and $\Sigma_R$ are each directed towards the forbidden region.
Eq.~(\ref{realHerring})
is the form given in \cite{wh} (a factor of $2$ appears because
we use a different normalisation of the Green's function). There are actually
four terms involved, depending on the combinations of directions in
which the derivatives in $\L$ and $\Lp$  act. Semiclassically, their 
superposition conspires to cancel all but trajectories that pass through
$\Sigma_L$ and $\Sigma_R$ in fixed senses, as we will see.

\subsection{The semiclassical Green's function}
\label{SCGsec}

\noindent The standard semiclassical approximation for the Green's function 
$G\argue$ takes the form of a sum over trajectories $\alpha$ which go
from $\x'$ to $\x$ with energy $E$,
\begin{equation}\label{realGreen}
	G\argue = \;\frac{1}{i\hbar} 
			\;\frac{1}{ (2\pi i\hbar)^{(\dim-1)/2} }
			\sum_\alpha
			    \sqrt{D_\alpha} \, 
				e^{iS_\alpha/\hbar}.
\end{equation}
For each orbit, $S_\alpha=\int_\alpha \p\cdot\d\x$ is the action and
\begin{equation}\label{giveD}
	D_\alpha = \frac{1}{\dot{x}\dot{x}'} \;
		 \det\left[-\dfdxdy{S_\alpha}{y}{y'}\right],
\end{equation}
assuming a single coordinate $x$ along the trajectory and $d-1$
coordinates $y$ transverse to it.
Detailed discussion can be found in \cite{gutz}. The usual presentation
of this formula has additional complex phases determined by the Maslov 
indices. In our case we assume that they are accounted for by the 
choice of branch for the square root of $D_\alpha$. 
We do it in this way so that complex trajectories can later be included
using the same formula, simply by complexifying the classical data 
entering into it.

In real WKB the sum is normally taken over all orbits $\alpha$ whose
time of evolution is positive. This corresponds to giving $E$ a small
positive imaginary part and results in the retarded Green's function
$G\argue$. We will also need its hermitean conjugate [see
Eq.~(\ref{GRLint})]. This is the advanced Green's function and can be
interpreted as a sum over all orbits whose time of evolution is
negative. Tunnelling effects are treated by including complex
trajectories \cite{BB}. This means that the initial and final
conditions in phase space can be complex, and that time varies over a 
contour in the complex plane, but the energy is still taken to be real. 
The complex trajectories contributing to the retarded and advanced
Green's functions are respectively those obtained by restricting the time
contour to the positive and negative half-planes.

The complete Green's function for the problem requires inclusion of
long trajectories, bouncing indefinitely in the wells. On the
contrary, we will restrict $\x$ and $\x'$ to the domain $\Vo$ and
include only trajectories which take a direct path from $\x'$ to $\x$
without leaving it. As long as caustic surfaces are avoided, which we
assume to be the case, this will result in a function which obeys the
defining equation for the Green's function while its arguments remain
in $\Vo$.  Thus it fits the description of the operator $\Gh$ used in
the abstract Green's identity Eq.~(\ref{absGreen}). The penalty for
the restriction on orbits is that the semiclassical approximation will
be badly behaved at caustics and will ultimately diverge exponentially
on the forbidden side of them. This will not concern us as long as the
caustics of the trajectories we use lie outside of $\Vo$. We need only
a local solution.  In particular, the elimination of multiple
reflections within a well means that $\Gh$ does not develop
singularities as the energy passes through energy levels.

\begin{figure}[h]
\vskip 0.3cm\hskip 2.5cm 
\psfig{figure=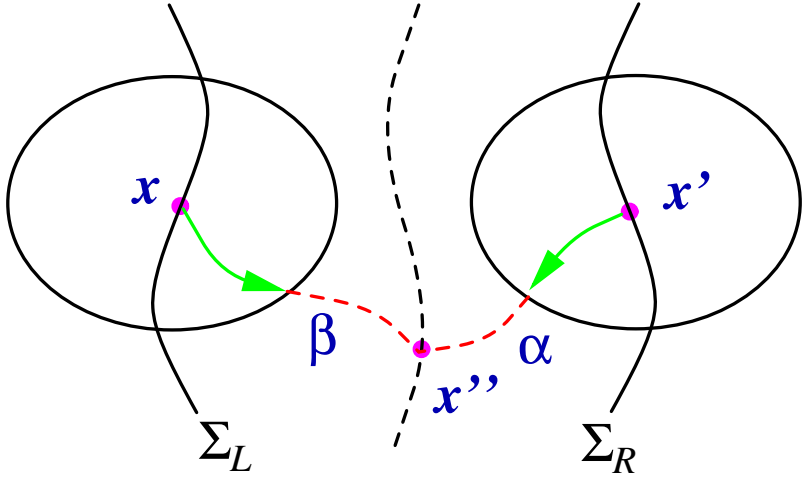,height=2.5in}
\vskip 0.5cm
\caption{\small A schematic illustration of the trajectories $\alpha$ and
$\beta$ respectively determining $G(\x'',\x',E)$ and
$G^\dagger(\x,\x'',E)$ in the integral defining $\Gt$. The arrows
indicate the velocity directions. Note that we expect that the
trajectories be generically complex, so that the paths also have
imaginary parts (not shown).}
\label{abfig}
\end{figure}

Our particular interest is to evaluate the integral in
Eq.~(\ref{GRLint}). This requires that we evaluate the retarded
Green's function connecting the initial position $\x'$ near $\Sigma_R$
to a point $\x''$ under the barrier and subsequently the advanced
Green's function from $\x''$ to the final position $\x$ near
$\Sigma_L$. We will assume, following the restrictions of trajectories
to $\Vo$ discussed above, that there is a single trajectory for each
leg of this journey, denoted respectively by $\alpha$ and $\beta$.
Consider $\alpha$. In a one-dimensional potential the trajectory
starts at $x'$ with velocity pointing towards the barrier and, 
after an evolution of time along the positive real axis, reaches 
a turning point.  A subsequent time evolution along 
the negative imaginary axis allows this orbit to penetrate the 
forbidden region with an action that has a positive imaginary part. 
For the second part $\beta$, time continues to evolve along the 
imaginary axis until the turning point on the left hand side is 
reached. Subsequent evolution of time parallel to the negative real 
axis (appropriate to the advanced Green's function)
allows the trajectory to reach $x$ with a final velocity pointing 
once again towards the barrier.  Notice that the
trajectory begins and ends at points which are symmetric in {\it phase
space}. It is easy to imagine a deformation of this picture into higher
dimensions, if $\x$, $\x''$ and $\x'$ are moved away from a symmetry
axis for example. The basic structure is sketched in
Fig.~\ref{abfig}.

To complete the calculation, a saddle point analysis of the integral
in Eq.~(\ref{GRLint}) is invoked, following substitution of these
complex orbits into Eq.~(\ref{realGreen}). The saddle point condition
specifies that the final momentum of $\alpha$ matches the initial
momentum of $\beta$, so that a concatenated trajectory
$\gamma=\beta\circ\alpha$ is defined going from $\x'$ to $\x$. It
turns out that the ensuing saddle point integration gives an
expression for $G_\tun\argue$ that is precisely of the form given in
Eq.~(\ref{realGreen}) except that the single orbit $\gamma$ is used
instead of the sum over $\alpha$.  This can be verified by explicit
calculation but it is a rather standard manipulation and we will not
go into it further here.

Before inserting $G_\tun\argue$ into Eq.~(\ref{realHerring}) and
writing out explicit results, it is helpful at this point to specify
more precisely the coordinates $\x=(x,y)$ that we will use. We assume
such a coordinate set to be constructed locally around each of the
sections $\Sigma_L$ and $\Sigma_R$ so that the positive $x$ direction
points towards the barrier and the area element is $\d s=\d^{\dim-1}y$. 
Under normal circumstances, when $\Sigma_L=\R\Sigma_R$, we will go 
further and assume the coordinate set around $\Sigma_L$ to be the 
symmetric image of the coordinate set around $\Sigma_R$. An important 
aspect of this convention is that, once a coordinate set is fixed on 
$\Sigma_R$, the orientation of the coordinate set on $\Sigma_L$ depends 
on the nature of $\R$. For example, in two dimensions
the frame on $\Sigma_L$ is oriented oppositely according 
to whether $\R$ is a reflection about an axis or an inversion through 
a point. This seemingly innocent difference will have an enormous 
effect on the nature of the final result. When calculated with respect 
to these frames, the matrix linearising motion about $\gamma$ is 
hyperbolic in the case of reflection symmetry and inverse hyperbolic 
in the case of inversion symmetry --- we will find as a direct 
consequence of this that the splittings are always positive 
in the former case whereas in the second they are sometimes 
negative. This will be discussed fully in section~\ref{positivity}.
Until then, however, our formulas will not depend overtly on
whether we have inversion or reflection as a symmetry as long as 
we keep to our conventions for $(x,y)$.

Now define the  following modification of $G_\tun\argue$,
\begin{equation} \label{defG}
	\Gtil(y,y',E) = \hbar\, \sqrt{\dot{x}_{\smash{\gamma}}
					     \dot{x}_{\smash{\gamma}}'} \;
			G_\tun(x_L(y),y,x_R(y'),y',E).
\end{equation}
We have scaled out by the velocities which will be reabsorbed into the
wavefunctions below; this mirrors the development in
\cite{bogomolnytrans}.  $\Gtil$ will always be evaluated with
arguments on $\Sigma_L$ and $\Sigma_R$, as appropriate, so the
$x$-dependences are determined by the implicitly-defined functions
$x_L(y)$ and $x_R(y')$, and can be suppressed.  Taking advantage of
the approximations, $-i\hbar\dydxh{G_\tun}{x} \approx \dot{x}_\gamma
G_\tun$ and $-i\hbar\dydxh{G_\tun}{x'}\approx-\dot{x}_\gamma' G_\tun$,
we may rewrite Eq.~(\ref{rlherring}) as follows,
\begin{equation}
	\Delta E  =  2\hbar \int_{\Sigma_L} \d y  \;
				\int_{\Sigma_R} \d y' \;\;
         \E(y,y') \; \Gtil(y,y',E) , 
\label{nicellHerring} 
\end{equation}
where,
\begin{equation}\label{defE}
	\E(y,y') =
\left[\frac{\left(\vhat_x  + \dot{x}^*_\gamma \right) 
				\psi_L(x_L(y),y)}
	         {2\sqrt{\dot{x}^*_{\smash{\gamma}}}}\right]^*
\left[\frac{\left(\vhat_x' + \dot{x}'_\gamma\right) 
				\psi_R(x_R(y'),y')}
                 {2\sqrt{\dot{x}_{\smash{\gamma}}'}}\right].
\end{equation}
Here we let $\vhat_x$ denote the projection 
of the velocity operator $-i\hbar\nabla$ onto the direction 
defined by the coordinate $x$.
If we ignore the fact that each of $\dot{x}_\gamma$ and
$\dot{x}'_\gamma$ depend on both initial and final position,
$\E(y,y')$ has the appearance of a product of wavefunctions,
\begin{equation}\label{decouplE}
	\E(y,y') = \psisr_\Ls (y)\, \psil_\Rs\!(y').
\end{equation}
Each of the wavefunctions here is meant to be identified with a
quantity within square brackets in the equation above.  The abuse of
notation inherent in Eq.~(\ref{decouplE}) is justified by the fact
that we will find that the dominant contribution to the integral comes
from a region of the $(y,y')$-plane small on classical scales, so that
$\dot{x}_\gamma$ and $\dot{x}'_\gamma$ can in practice be replaced by
constants.

The arrows on the wavefunctions indicate directions along which flux
is predominantly flowing, which arises as follows.  The surface
$\Sigma_R$ defines two surfaces of section in phase space, $\Sigr_R$
and $\Sigl_R$, corresponding to the two senses of $\dot{x}$. If we
look at a phase space representation such as the Wigner function, the
combination of $\dot{x}'_\gamma$ and the velocity operator in
Eq.~(\ref{defE}) leads to enhancement of $\psil_\Rs$ around $\Sigl_R$
and suppression around $\Sigr_R$ (and equivalently for
$\psir_\Ls$). Therefore we can take the arrows to indicate left- and
right-going wavefunctions.  The particular combination seen in
Eq.~(\ref{decouplE}) arose because we chose retarded Green's functions
to extend both of the states into the forbidden region.  Had we chosen
other combinations of advanced and retarded Green's functions,
$\dot{x}_\gamma$ and $\dot{x}'_\gamma$, would have entered
Eq.~(\ref{defE}) with different signs, leading to different
combinations of left and rightgoing wavefunctions.  It is clearly
preferable that this choice be made so that there is symmetry between
the initial and final coordinates. In particular, once we have chosen
between advanced and retarded Green's functions for wavefunction
continuation, the same choice is applied to both wavefunctions in the
matrix element for $\Delta E$. In phase space, this means that the
splitting is calculated from the behaviour of states either near
$\Sigl_L$ and $\Sigr_R=\R\!\Sigl_L$, or else near $\Sigr_L$ and
$\Sigl_R=\R\!\Sigr_L$.

\begin{figure}[h]
\vskip 0.cm\hskip 3.5cm 
\psfig{figure=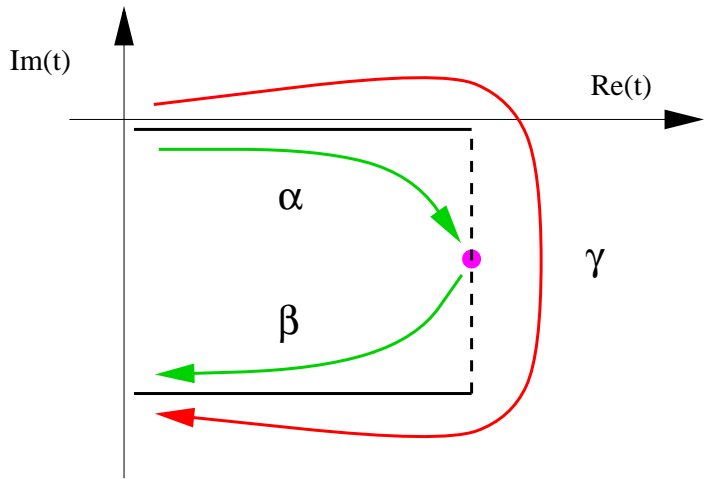,height=2.5in}
\caption{\small The paths taken in the complex time plane by the segments
$\alpha$ and $\beta$, and their concatenation $\gamma$. Deformation 
of these contours is allowed as long as branch points are avoided.}
\label{tpfig}
\end{figure}

A consequence of this symmetry is that, taking into account the
complex conjugation of $\psi_L(\x)$ in the matrix element, the time
evolution of $\gamma$ is directed along opposite directions of the
real axis before and after barrier penetration (Fig.~\ref{tpfig}).
Therefore there is cancellation in the real time along the path
$\gamma$ and the net time of evolution is directed predominantly along
the negative imaginary axis. For the special case of the periodic
bounce orbit $\gamma_0$, this cancellation is complete and the total
time $-i\tau_0$ is pure imaginary. In computing these trajectories, it
is tempting to deform the time contour so that it flows directly down
the imaginary axis, forgoing the real time segments.  In general that
is dangerous, however, unless one can guarantee that branch points are
not crossed as the contour is deformed. Besides, the contour defined
above gives trajectories whose position coordinates remain
predominantly real throughout the evolution, which has obvious
intuitive advantages.

To reflect the near-imaginary time evolution, let us use the notation,
\begin{equation} \label{defK}
                 K(y,y',E) = -iS_{\gamma}(x_L(y),y,x_R(y'),y',E)
\end{equation}
to denote the ``Euclidean'' action of the path $\gamma$, suppressing $x$ as
usual. As with the velocities $\dot{x}$ and $\dot{x}'$, the quantity
$K$ is predominantly positive but is somewhat complex in general,
implying that $S$ is dominantly positive imaginary but with some
real component. An explicit expression for the modified Green's
function is then,
\begin{equation}\label{RLGreen}
	\Gtil(y,y',E)  = \;
			\;\frac{1}{ (2\pi\hbar)^{(\dim-1)/2} } \;
	\sqrt{ \chi_\R\det \!-\!\dfdxdy{K}{y}{y'} } \; e^{-K/\hbar}.
\end{equation}
Note that this has a different $\hbar$-dependence than
(\ref{realGreen}) and also has no velocity prefactors.
The sign $\chi_\R$ is the same as appeared in Eq.~(\ref{trunc})
and accounts for any reorientation of the local $y$ 
coordinates relative to the global cartesian coordinates
with which we started.
Except for the fact that it uses complex trajectories, the construction 
of $\Gtil(y,y',E)$ is precisely analogous to that of the 
Bogomolny transfer operator \cite{bogomolnytrans}. Likewise, the 
arrowed wavefunctions can be interpreted as states in the related, 
semiclassically-defined Hilbert space. This interpretation will 
be developed further in section~\ref{Transop}.

It is at this point that our presentation differs most markedly from
previous developments. The discussion leading to Eq.~(\ref{RLGreen})
is invalid when the end points of $\gamma$ are too close to the energy
contour $E=V$. In \cite{wh} a careful treatment of this problem is
given in which a uniformised semiclassical treatment of the Green's
function, using Airy functions, is developed. Auerbach and Kivelson
\cite{AK} choose to avoid the turning point singularities by taking
the surface to be well inside the forbidden region. Our approach is
similarly to assume $\Sigma_L$ and $\Sigma_R$ to be well separated
from the energy contours around the dominant trajectories, but on the
allowed side. In particular, this means that we must avoid the natural
temptation to continue the wavefunctions from as close to the barrier
as possible, but does have the advantage that the ensuing calculation
is much simpler (later we will also appeal to symplectic invariance to argue
that it should be possible to compute in a Poincar\'e section
symplectically rotated in phase space so that it remains well-defined
even at the turning point). It is also advantageous only to consider
the wavefunction in the allowed region and to put all exponentially
small quantities and the attendant complex structure into the
Green's function and quantities derived from it.

As a function of $y'$ and $y$, the real part of $K(y,y',E)$, which
controls the magnitude of $\Gtil(y,y',E)$, is minimised 
when the orbit intersects the Poincar\'e sections with real momenta.
This condition is fulfilled by the periodic
bounce orbit $\gamma_0$, for which the net time of evolution is 
imaginary and $K$ is real. Recall that this is also the orbit from 
which $f_0$ is
determined in periodic orbit theory. It can be argued quite generally
that the segments of this trajectory corresponding to real time
evolution take place in real phase space, connecting to a tunnelling
segment of imaginary time evolution and complex coordinates at a
turning point where $\dot{\x}=0$ if there is time-reversal symmetry,
or $\Im\,\dot{\x}=0$ more generally \cite{dtun,therussians}. In the potentials
treated here, the bounce orbit is on a symmetry axis for which
$y=y'=0$. When its arguments move away from $\gamma_0$, the Green's 
function then decays in magnitude over a length scale of 
$O(\hbar^{1/2})$.

We will assume that the dominant contribution to the splitting matrix
element comes from this neighbourhood. Approximating the amplitude of
the Green's function by its value for $\gamma_0$,
Eq.~(\ref{nicellHerring}) becomes
\begin{equation}\label{GHerring}
	\Delta E = 2\hbar\, \Gtil(y_0,y_0,E) \int_{\Sigma_L} \d y \;
		\int_{\Sigma_R}\d y' \;\;
           \E(y,y')\; e^{-\K2(y,y')/\hbar}
\end{equation}
where $\K2(y,y')$ is the quadratic form giving the second variation of
$K\argue$ about the intersection of $\gamma_0$ with the sections (for
which the coordinates are denoted $y_0$). At the same level of
approximation we may choose $\E(y,y')$ to have the decoupled form
(\ref{decouplE}) by taking $\dot{x}$ and $\dot{x}'$ to be the values
for the central orbit $\gamma_0$. $\K2(y,y')$ is calculated from the
complex monodromy matrix $W$ defined by $\gamma_0$, which has the
following block form [each block having dimension
$(\dim-1)\times(\dim-1)$],
\begin{equation}\label{defW}
	     W = \left( \matrix { \phantom{i}w^\dagger &           -iu \cr 
		                                    iv & \phantom{-i}w  }  
							\right), 
\end{equation}
where $u$ and $v$ are hermitean (real if $d=2$). That $W$ is of this 
form follows from symplecticity  and the property
\begin{equation}\label{revcon}
		W^{-1} = W^*,
\end{equation}
which itself is a consequence of the fact that taking the complex 
conjugate of the path of $\gamma_0$ in phase space gives its time 
reversed partner, up to the symmetry operation $\sigma$. The form
in Eq.~(\ref{defW}) is less restricted than in  Eq.~(\ref{defW0})
because $\gamma_0$ includes real orbit segments at either end.
Standard manipulation using generating function conditions obeyed by
the action now gives,
\begin{equation} \label{defKmat}
%		\K2(y,y') =\frac{1}{2}\spinor{y}{y'}^T
		\K2(y,y') =\frac{1}{2}\left(y \;\; y'\right)
	         \left( \matrix { a & b^*  \cr 
		                  b & a^*  }  
							\right) 
		\spinor{y}{y'}   ,
\end{equation}
where,
\begin{equation}
	a =  wu^{-1} \qquad\mbox{and}\qquad b =  -u^{-1}
\end{equation}
are respectively symmetric and hermitean.  We note that $y$ and $y'$
are in general $(d-1)$-dimensional and the various sub-matrices are
also of dimension $d-1$. In the case $d=2$ they are 
scalars. Notice that, as an operator kernel, $\K2(y,y')$ is formally
hermitean (it is only formally so because we have not yet specified
the Hilbert space). This hermiticity is a consequence of the inversion
symmetry underlying Eq.~(\ref{revcon}) --- {\it i.e.}  that complex
conjugation amounts to time reversal --- and is not confined to the
quadratic truncation.  To aid practical implementation, we note that
the amplitude in Eq.~(\ref{RLGreen}) is
$\det\!\left[-\dfdxdyh{K}{y}{y'}\right] = \det{u}^{-1}$.

It is important to note that the integration in Eq.~(\ref{GHerring})
must be treated with care, since the wavefunction will typically vary
significantly on the length scale over which the kernel of this matrix
element decays. In particular, if we were to implement some sort of
WKB approximation of the wavefunctions (such as a collective
approximation using Green's functions, for example), the center of our
expansion would not be a saddle point of the whole integrand ---
further saddle point analysis would require that we revert to the full
expression for $\Gtil(y,y',E)$ in that case. In order for
the quadratic phase expansion to be valid, the integration must be
interpreted as an exact sum. Even with this limitation,
Eq.~(\ref{GHerring}) is useful because it can in practice be evaluated
following a relatively painless numerical diagonalisation to get the
wave function, as we will see.

In any case, we have reduced calculation of the splitting to
evaluation of a simple matrix element. Either
Eq.~(\ref{nicellHerring}) or its approximation (\ref{GHerring}) could
now be used to determine the splittings semiclassically. However, we
choose to postpone this step because there remains additional
structure which has not yet been addressed. In particular, there is an
appealing canonical invariance which becomes manifest when the theory
is reformulated in a phase space representation, as we now discuss.

\section{A Phase Space Formulation and Scarring}
\label{apsfas}

\noindent The matrix element for $\Delta E$ developed in the previous
section is conveniently interpreted in the Weyl formalism. The Weyl
symbol of any operator $\O$ is defined by,
\begin{equation}\label{defWeyl}
	W_\O(\q,\p) = \int \d \s \;  e^{i\p\cdot\s/\hbar} \;
		\braopket{\q+\s/2}{\O}{\q-\s/2}.
\end{equation}
and for a state $\ket{\psi}$ the Wigner function $W_\psi(\q,\p)$
is defined by substituting $\O=\ket{\psi}\bra{\psi}/(2\pi\hbar)^\dim$. 
The expectation value of $\O$ can then be calculated as
follows,
\begin{eqnarray}\label{Weylmat}
	\braopket{\psi}{\O}{\psi} 
		= \int\d\q\d\p\; W_\psi(\q,\p) W_\O(\q,\p).
\end{eqnarray}
We would like to do the same for the matrix element in 
Eq.~(\ref{GHerring}).

Because we deal with integrations which are restricted to sections
(which we will assume to be defined by fixing $x$) it is convenient to
define the partial Weyl symbol
\begin{equation}\label{halfassedWeyl}
		 \Wt_{\Gtil}(\zetab,E) = \int \d s \;  e^{ip_y s/\hbar} \;
	         \Gtil (y+s/2,y-s/2,E)
\end{equation}
and the partial Wigner function
\begin{equation}\label{halfassedWigner}
		 \Wl_n\!(\zetab) = 2\pi\hbar \int \d s \;  e^{ip_y s/\hbar} \;
		 \E(y+s/2,y-s/2),
\end{equation}
where we denote,
\begin{equation}\label{spinor}
	\zetab = \spinor{y}{p_y}.
\end{equation}
We have taken the opportunity here to reintroduce the index $n$
labelling the different doublets. We will often refer to
$\Wl_n\!(\zetab)$ as the Poincar\'e-Wigner function, which we
interpret as a function on the {\it oriented} section
$\Sigl_R$. Canonically invariant expressions will be given
below for $\Wt_{\Gtil}$.

These functions will usually be evaluated with $x$ and $x'$ restricted to
sections and we therefore suppress these variables.  These two
quantities are then thought of as functions of the section coordinates
$y$ and $p_y$. In general these are of dimension $\dim-1$, in
which case the $s$ integral is of the same dimension  and the term
$ip_y s$ in the exponential should be treated as a dot product. Notice
that the choice of normalisation for $\Wl_n\!(\zetab)$ is not
conventional. This will simplify the appearance of the final
result. It should be remarked that the partial Weyl function could be
defined for any operator defined (at least semiclassically) on the
Poincar\'e section by replacing $\Gtil$ with the operator of
interest. Similarly we can replace $\E(y,y')$ by $\psi(y)\psi^*(y')$
if we are interested in the properties of a semiclassically defined
wavefunction $\psi(y)$ defined on the section. Finally, we note that
these transforms are strictly speaking properly-defined only when
$(x,y)$ are linear, cartesian coordinates, so that the sections
$\Sigma_R$ and $\Sigma_L$ are planes.  However, since we will always
confine ourselves to semiclassical approximation, and usually to
integrations confined to the immediate neighbourhood of the orbit
$\gamma$, it should not be a problem in practice to work with
curvilinear sections.

In analogy with Eq.~(\ref{Weylmat}), the integrations over $\Sigma_R$
in Eq.~(\ref{nicellHerring}) can be converted to a phase space
integral on the section using the partial symbols as follows,
\begin{equation}\label{WHerring}
	        \Delta E_n = {1\over\pi} \int_{\Siglsub_R} 
		\frac{\d\zetab}{(2\pi\hbar)^{\dim-1}}\;\;
		\Wt_{\Gtil}(\zetab,E_n)\;\Wl_n\!(\zetab).
\end{equation}
To evaluate this we need an explicit expression for 
$\Wt_{\Gtil}(\zetab,E)$.

$\Wt_{\Gtil}(\zetab,E)$ is structurally similar to the standard Weyl
symbol of a time propagator. An explicit semiclassical expression for
the latter was developed in \cite{berryscars}. Since the detailed
calculation is easily transcribed from that case, we will just write
down the results. For a given point $\zetab$ on $\Sigl_R$,
$\Wt_{\Gtil}(\zetab,E)$ is constructed using the ``midpoint orbit''.
This is the (assumed unique) orbit of type $\gamma=\beta\circ\alpha$
going from $\zetab_A$ on $\Sigl_R$ to $\R\zetab_B$ on $\Sigr_L$ such
that the midpoint is $\zetab = (\zetab_A+\zetab_B)/2$.  We associate
with it the complex action,
\begin{equation}\label{area}
	\A(\zetab,E) = K(\x_B,\x_A,E) + ip_y(y_B-y_A).
\end{equation}
The contribution coming from the $y$ degree of freedom can be interpreted 
as the symplectic area in the $y$-$p_y$ plane of a chord defined by the 
trajectory and the straight line connecting initial and final points. 
This is invariant under linear canonical transformations in $(y,p_y)$.  
Geometrical interpretation is more awkward for the full expression, but 
canonical invariance within a surface of section is certainly maintained.  
We can now write \cite{berryscars},
\begin{equation}\label{midpoint}
	\Wt_{\Gtil} (\zetab,E) 
			= \frac{e^{-\A(\zetabs,E)/\hbar}}
		{\sqrt{\chi_\R\det\left[(W_{AB}+I) / 2 \right]} }
\end{equation}
where $W_{AB}$ is the complex symplectic matrix linearising motion
about the midpoint orbit in $(y,p_y)$ coordinates and $\chi_\sigma$ 
is the usual sign compensating for any reorientation of the local 
coordinates $y$ with respect to the global ones.

The Gaussian approximation employed in Eq.~(\ref{GHerring})
corresponds to replacing $W_{AB}$ by its value $W$ on $\gamma_0$
and using the following quadratic approximation for $\A$,
\begin{equation}\label{qarea}
	\A(\zetab,E)-K_0 \approx -i \zetab^T J \frac{W-I}{W+I} \zetab
		   = -i \Omega\left(\zetab,\frac{W-I}{W+I}\zetab\right)
\end{equation}
where $T$ denotes transpose and $\Omega(\xi,\eta)=\xi^T J\eta$ is the
symplectic two-form, $J$ being the unit symplectic matrix. This result
is derived starting with $\zetab_B=W\zetab_A$ and employing some
generating function conditions satisfied by $K(\x_B,\x_A,E)$ to compute the 
second derivatives of
$\A$. A more complete discussion can be found in \cite{berryscars}
(beware, however, that $J$ is defined with the opposite sign there).

For us the feature of most immediate interest is that $\A$ is real for
real $(y,p_y)$.  This follows from the property $W^{-1} = W^*$
discussed earlier.  This inversion symmetry implies that the matrix
\begin{equation}
     \left(\frac{W-I}{W+I}\right)^* = \frac{W^{-1}-I}{W^{-1}+I}
				    = -\frac{W-I}{W+I}
\end{equation}
is imaginary, which in turn, taking account of the factor of $i$ in
Eq.~(\ref{qarea}), implies that $\A(\zetab,E)$ is real to quadratic
order.  We are therefore free to write
\begin{equation}\label{area'}
	\A(\zetab,E) \approx K_0 +  \zetab^T M \zetab
\end{equation}
where the matrix
\begin{equation} \label{Mdef}
           M=-iJ\frac{W-I}{W+I}
\end{equation}
is real, symmetric and, it turns out, positive definite.

\begin{figure}[h]
\vspace*{0.5cm}\hskip 2.0cm 
\psfig{figure=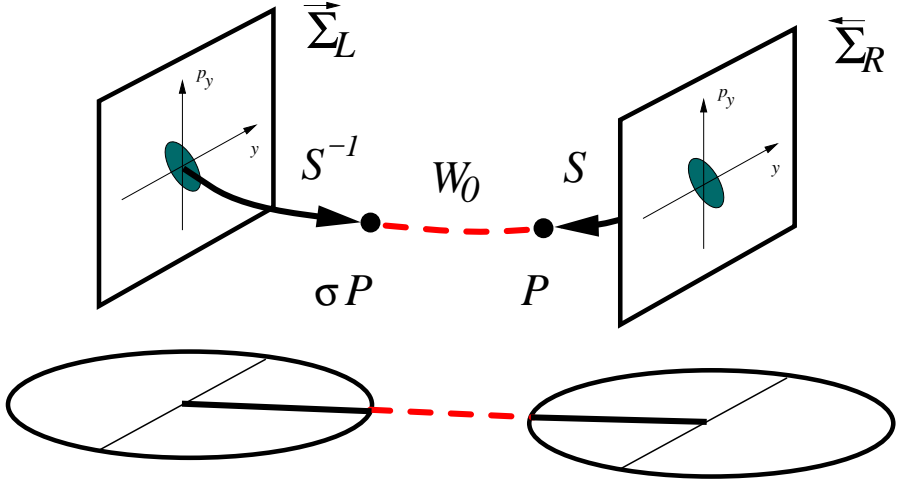,height=2.5in}
\vspace*{0.5cm}
\caption{\small A schematic representation of the structures needed to
calculate the splitting. First the imaginary-time trajectory 
connecting the surfaces of section on each side of the barrier
must be computed. This defines a complex monodromy matrix of 
the form $W=S^{-1} W_0 S$ where $S$ is the real matrix
linearising motion around a real trajectory segment. $W$ defines a 
Gaussian $g_M(\zetab)$, approximately supported in an elliptical domain
of area $O(\hbar)$ in each Poincar\'e section, as indicated
by the shaded regions. Overlap with the Poincar\'e-Wigner function of 
the state in this region gives the splitting.}
\label{SOSfig}
\end{figure}

If we factor out the function $f_0(E)$ representing the mean splitting, 
the Gaussian approximation to the symbol $\Wt_{\Gtil} (\zetab,E)$
becomes
\begin{equation} \Wt_{\Gtil} (\zetab,E) \approx
		\pi(2\pi\hbar)^{d-1} f_0 (E) g_M(\zetab),
\end{equation}
where the Gaussian 
\begin{equation}\label{gaussM}
	g_M(\zetab) = \frac{\sqrt{\det M}}{(\pi\hbar)^{\dim-1}}
		\; e^{-\zetabs^T M \zetabs/\hbar}
\end{equation}
is normalised so that $\int\d\zetab\; g_M(\zetab)=1$.  Note that we
have used the fact that $W$ is related to the matrix $W_0$ of
section~\ref{theoryforthemean} by the similarity transformation
$W=S^{-1}W_0S$ where $S$ is the real matrix linearising motion from
$\Sigl_R$ to the barrier (see below and Fig.~\ref{SOSfig}).

The splitting (\ref{WHerring}) can then be put in the form,
\begin{equation} \label{enfin}
	\Delta E_n  =  f_0(E_n){\displaystyle\int_{\Siglsub_R}}\d\zetab\;
			  g_M(\zetab) \Wl_n\!(\zetab).
\end{equation}
Alternatively, we can write
\begin{equation}
	y_n = \rho_0(E_n){\displaystyle\int_{\Siglsub_R}}\d\zetab\;
			g_M(\zetab) \Wl_n\!(\zetab) 
\label{finally}
\end{equation}
for the rescaled splitting defined by $y_n = \rho_0(E_n)\Delta E_n/ f_0(E_n)$ 
in section~\ref{theoryforthemean}.

Eq.~(\ref{finally}) is the central theoretical result of the
paper. Even though the calculation leading to it may seem complicated,
this final form is very simply interpreted and implemented. Let us
summarise the ingredients needed to use it. The first step is to find the
bounce orbit which crosses the barrier in imaginary time and is
responsible for the contribution $f_0(E)$ to Eq.~(\ref{deff}). This
reduces to a trivial one-dimensional problem in the common case that
it lies in a symmetry axis, but is a straightforward numerical task
even in the most general case. Quite generally this bounce orbit
has a turning point $P$ where its coordinates in phase space are
real. Take $P$ and its symmetric image $\R P$ to be the initial and
final points of the imaginary-time trajectory, denoted
$\gamma_B$. Starting at $P$, one can equally consider evolution in
{\it real} time, defining a real trajectory $\gamma_R$ going in
negative time to $\Sigl_R$.  Its image $\gamma_L=\R\gamma_R$ goes to
$\R P$ from $\Sigr_L$.  Then the concatenation
$\gamma_0=\gamma_L^*\circ\gamma_B\circ\gamma_R$, where the star denotes
time reversal, is an orbit going from $\Sigl_R$ to $\Sigr_L$ in a net
time which is imaginary. The monodromy matrix $W=S^{-1}W_0S$ of
$\gamma_0$ defines $M$ in Eq.~(\ref{gaussM}). The final ingredient is
the computation of the Poincar\'e-Wigner function $\Wl_n\!(\zeta)$. There
are no simple semiclassical theories in the case that the dynamics is
chaotic. It can be calculated numerically however, with the advantage
that the precision need not be excessively high. One can also model
the wave function from random matrix ensembles, and
Eq.~(\ref{finally}) will form the basis for quantitative statistical
analysis of the splittings in that case.

In cases of additional symmetry, one encounters the situation where
the real extension to $\gamma_B$ is a periodic orbit.  In this situation 
it is common to interpret an exceptionally large accumulation of probability
amplitude in its neighbourhood as a manifestation of scarring
\cite{hellerscars,berryscars}. Our expression for the  $y_n$ is
in such cases simultaneously a ``scar-ometer'' --- it gives a precise
measure of the degree to which the state is scarred by the periodic
orbit.  In fact, it is extremely close to being the value of a Husimi
distribution evaluated on the orbit. It would be precisely so if
$g_M(\zetab)$ were the Wigner function of a coherent state. It is not
--- to be a Wigner function of a pure state requires $\det M = 1$,
which is not true in our case --- but it is very close. In particular,
a proper Husimi distribution would take the same form as
Eq.~(\ref{finally}) except for a rescaling of the matrix $M$. The
other difference between our result and a standard Husimi functon is
that our overlap is confined to a lower-dimensional Poincar\'e
section. The fact that there is a strong correlation between scarring
and the tunnelling rate in such cases is not at all surprising.
Scarred states have a large accumulation of probability along the
optimal tunnelling route defined by $\gamma_0$ and would be expected
to tunnel at a higher rate. The fact that this observation can be
given a precise quantification is nontrivial, however.

As a scar-ometer alone, the expression has the following interesting,
and potentially useful property: our calculation indicates, and
explicit numerical calulation in section~\ref{thesystem} will verify,
that the overlap is an invariant of the orbit. That is, it does not
matter for which section $\Sigma_R$ ({\it i.e.} which value of $x$) we
evaluate the overlap, the result is always the same within
semiclassical error. It gives a single, unambiguous number with which
to quantify the degree of scarring of the state.  An important element
in this invariance is that the elliptic region defined in the
Poincar\'e section by level curves of $\zetab^TM\zetab$ deforms with
the flow as we move from one point of the orbit to another --- as a
consequence of $W=S^{-1}W_0S$ we have $M=S^T M_0 S$, where $S$
linearises the flow along the real orbit. In particular, as we move
further from the turning point at which tunnelling begins, this
elliptic domain becomes stretched along the stable manifold of the
periodic orbit.  Clearly, if the evolution is taken too far, errors of
linearised propagation will grow and the invariance will fail. However
for moderate deformations invariance is expected and observed.

Finally, we would like to verify that the result is correctly
normalised. To this end we note that the general property of standard
Wigner functions,
\begin{equation}\label{MCW}
	\left\langle \sum_n W_n(\q,\p)\;\delta(E-E_n) \right\rangle 
				= \frac{\delta(H(\q,\p)-E)}{(2\pi\hbar)^\dim}
\end{equation}
can be argued (appendix~\ref{TFPW}) to have the following analog in the 
case of Poincar\'e-Wigner functions;
\begin{equation}\label{MCPW}
	\left\langle \sum_n \Wl_n\!(\zetab)\;\delta(E-E_n) \right\rangle 
				= \chi_E(\zetab).
\end{equation}	
Here $\chi_E(\zetab)$ is the characteristic function of the
energetically-allowed region in the $\zetab$-plane ({\it i.e.}
$\chi_E(\zetab)$ is unity whenever $\zetab$ is within the
energetically permitted domain and is zero otherwise.)  The average
here is meant to be taken over an energy range small on classical
scales but large enough to smooth out quantum fluctuations.  This
tells us that the Poincar\'e-Wigner function is expected on average to
be spread uniformly over the surface of section in the canonical
measure $\d y\d p_y$, which is unsurprising (note however that
renormalistion of the wave function by the velocity as in
Eq.(\ref{defE}) is a necessary ingredient for this to work). A fuller
discussion of this, along with certain qualifications, is offered in
appendix~\ref{TFPW}. As a consequence,
\begin{equation}
	\left\langle\sum_n y_n\delta(E-E_n) \right\rangle = \rho_0(E).
\end{equation}	
which follows from taking the average inside the $\zetab$ integral of
Eq.~(\ref{finally}) and taking $g_M(\zetab)$ out.  In other words,
$\langle y_n \rangle=1$ as required. 

We note finally that fluctuations of $\sum_n
\Wl_n\!(\zetab)\;\delta(E-E_n)$ about the mean defined above can be
treated semiclassically as a sum over real midpoint orbits returning
to $\Sigl_R$ \cite{berryscars}. We can combine these sums with our
matrix element for $y_n$ to derive semiclassical expressions for the
fluctuations of $\sum y_n \delta(E-E_n)$.  In doing so we simply
reproduce the expression for $f(E)$ as a sum over complex orbits which
was found using the trace formula \cite{us}. Though less direct, this
derivation has technical advantages because the long orbits can be
effectively treated within real dynamics and ambiguities due to
the possible intervention of  Stokes phenomena etc, normally difficult 
to control in higher dimensions, disappear.

\section{The Case of Resonances}
\label{resonancescase}

\noindent
Now we outline how the discussion of the previous sections may be
adapted to the calculation of resonant lifetimes in metastable
systems. Many of the formal expressions in this case look
identical. There are important differences in detail, however,
particularly in the boundary conditions placed on the imaginary-period
bounce orbit which mediates tunnelling.

\begin{figure}[h]
\vspace*{0.5cm}\hskip 3.0cm 
\psfig{figure=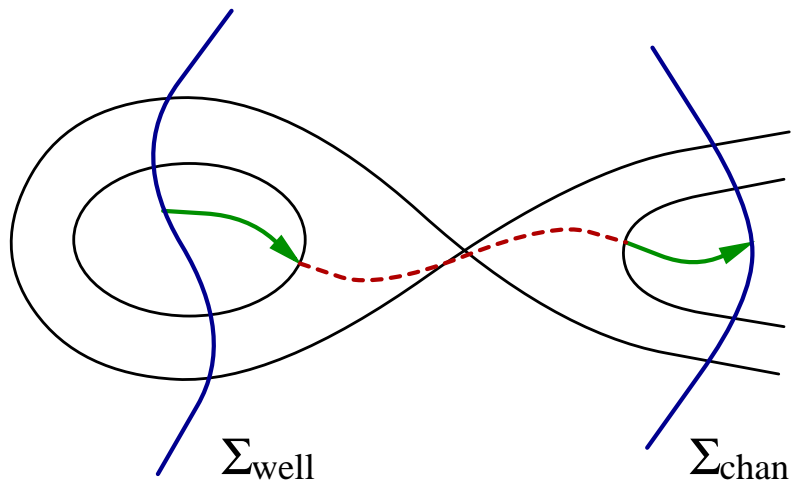,height=2.5in}
\vspace*{0.5cm}
\caption{\small Potential contours of a  typical metastable system. 
The section $\Sigma_{\rm chan}$ is defined so that, restricting
integration to the left of it, the resonant wavefunction is
normalised. The resonance width is related to the flux across it. The
wavefunction is assumed known on the section $\Sigma_{\rm well}$
cutting through the well and a semiclassical Green's function is used
to propagate out to $\Sigma_{\rm chan}$.  The orbit shown is
appropriate to the retarded Green's function and corresponds to type
$\alpha$.}
\label{resonancefig}
\end{figure}

To be concrete, we will initially assume a potential with a fairly
restricted topography, to be relaxed later.  As illustrated in
Fig.~\ref{resonancefig}, we assume a single, classically bound
potential well, with a barrier leading to a single escape channel. In
quantum mechanics, we find approximate states localised within the
well. These are associated with poles of the scattering matrix at
complex energies
\begin{equation}\label{defgamma}
	E_n = E_n^0 - \frac{i\Gamma_n}{2}
\end{equation}
where $\Gamma_n$ is the exponentially small resonance width which is
the object of discussion in this section. (Note that there will also
typically be resonances with very large widths associated with
resonance wavefunctions which have negligible amplitude inside the
well. Their widths are not mediated by tunnelling so such states do
not concern us here.)

Letting $\ket{\psi_n}$ be the metastable state associated with a
particular resonance, we formally write the Schr\"odinger equation,
\begin{equation}\label{resSchrod}
	\Hh\ket{\psi_n} = E_n^0\ket{\psi_n} - \frac{i\Gamma_n}{2}\ket{\psi_n}.
\end{equation}
Being a metastable state, $\ket{\psi_n}$ is unnormalised, but it is
not difficult to define an operator $\Thhat$ so that
$\Thhat\ket{\psi_n}$ is both normalised and an approximate
eigenstate. Typically, we contruct some region $A$ enclosing the well
and excluding all but finite portions of any escape channels and let
$\Thhat$ represent multiplication of the wavefunction by its
characteristic function. For example, in Fig.~\ref{resonancefig} we
might let $A$ be the region to the left of $\Sigma_{\rm chan}$. We
will further assume $\ket{\psi_n}$ to be normalised so that
$\braopket{\psi_n}{\Thhat}{\psi_n}\approx 1$. Applying $\Thhat$ to the
equation above, taking matrix elements and then subtracting the
complex conjugate, we get the following modification of the abstract
Herring formula,
\begin{equation}\label{GammaHerring}
	\Gamma_n = i\braopket{\psi_n}{\left[\Thhat,H\right]}{\psi_n}
		= i\langle\psi_n
		\dbar_{\Sigma_{\rm chan}}
		\psi_n\rangle.
\end{equation}
Physically, this relates the resonance width to the flux leaving the well.

As before, we use a semiclassical Green's function to extend the
wavefunction from the interior of the well across the barrier and into
the escape channel, allowing evaluation of Eq.~(\ref{GammaHerring})
along $\Sigma_{\rm chan}$. We assume a section $\Sigma_{\rm well}$ can
be defined in the well such that a single complex trajectory connects
it with $\Sigma_{\rm chan}$ while remaining free of caustics in the
region enclosed between them (or at least this should be so where the
dominant contributions to integration arise). In other words, this
single trajectory defines semiclassically a Green's function $\Gh$
which satisfies Schr\"odinger's equation in the region where Green's
theorem demands it. Such a trajectory will be labelled by $\alpha$ and
is illustrated in Fig.~\ref{resonancefig}. Notice that $\alpha$
completely crosses the barrier and has real-time segments before and
after the crossing. Thus it looks superficially like the orbit
$\gamma$ of the previous section. The important difference is that, in
the typical case where we use the retarded Green's function, the real
time evolution of $\alpha$ is along the postive axis before {\it and }
after the crossing.

These real parts will be cancelled when we integrate the Green's
function against its hermitean conjugate, following evaluation of the
diagonal matrix element in Eq.~(\ref{GammaHerring}). To see this more
explicitly, we note that calculation of the resonance width using the
sequence of the previous section leads us to define the following
tunnelling operator,
\begin{equation}\label{defGtun}
	\Gt = % i\Ghd\left[\Thhat,\Hh\right]\Gh,
	 i\Ghd\dbar_{\Sigma_{\rm chan}}\Gh,
\end{equation}
in terms of which the resonance width is,
\begin{equation}
	\Gamma_n  = 	
	       -\langle\psi_n
		\dbar_{\Sigma_{\rm well}}
			\Gt
		\dbar_{\Sigma_{\rm well}}
		\psi_n\rangle.
\end{equation}
The sectional overlap restricts integration to the neighbourhood of
$\Sigma_{\rm well}$ and is defined in obvious analogy with the
previous sections.  Except for the factor of $i$, $\Gt$ looks formally
identical to its splitting counterpart defined in
Eq.~(\ref{concatGdG}). As in that case, the real time evolutions
coming from $\Gh$ and $\Ghd$ will cancel --- following saddle point
integration, $\Gt$ is approximated with a single orbit
$\gamma=\beta\circ\alpha$ obtained by concatenating an orbit $\alpha$
of positive real time evolution with an orbit $\beta$ of negative real
time evolution.

Despite its formal similarity with that of the previous sections, it
is important to stress that the single orbit $\gamma$ from which the
present $\Gt$ is constructed has very different boundary conditions
imposed upon it. In particular, it crosses the barrier {\it twice},
beginning and ending on $\Sigma_{\rm well}$. It will be obtained by
deforming an imaginary time orbit which is periodic, not
pseudoperiodic as in the previous section (in any case, there is no
longer a symmetry with respect to which pseudoperiodicity could be
defined).

Following detailed pursuit of this program, we arrive at the
following Poincar\'e matrix element  for $\Gamma_n$,
\begin{equation}
	\Gamma_n  =  \hbar \int_{\Sigma_{\rm well}} \d y  \;
				\int_{\Sigma_{\rm well}} \d y' \;\;
        \Gtil(y,y',E) \; \E_n(y,y'), 
\end{equation}
where $\E_n(y,y')$ is defined analogously to Eq.~(\ref{defE}). The
Poincar\'e Green's function $\Gtil(y,y',E)$ takes exactly the same
form as given in Eq.~(\ref{RLGreen}), except that the imaginary action
$K(y,y',E)$ is defined by the double-bounce orbit appropriate to
resonances as described above. Note that this is in spite of the
factor $i$ in Eq.~(\ref{defGtun}), which is compensated by different
phases which arise on passing through the barrier on the way to
$\Sigma_{\rm chan}$, as can be verified in standard one-dimensional
WKB.

\begin{figure}[h]
\vspace*{0.5cm}\hskip 3.0cm 
\psfig{figure=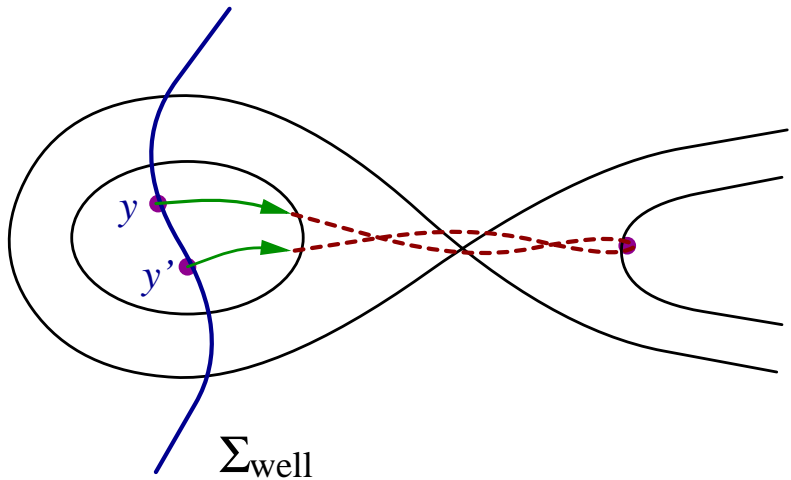,height=2.5in}
\vspace*{0.5cm}
\caption{\small For resonances, the orbit $\gamma$ takes the path shown. As 
usual, the arrows show the velocity direction during the real time
segments.}
\label{dgammafig}
\end{figure}

Having performed the saddle point manipulation leading to $\Gtil
(y,y',E)$, the real time segments connecting the outer turning point
of imaginary evolution to $\Sigma_{\rm chan}$ can be dispensed with,
and in fact $\Sigma_{\rm chan}$ disappears from the discussion, just
as the intermediate section did in the previous sections.  Beforehand,
however, the detailed saddle point analysis is more straightforward if
$\Sigma_{\rm chan}$ is chosen to the right of the barrier. If it is
taken under the barrier as in the previous section, the requirements
vplaced on $\Thhat$ leading to Eq.~(\ref{GammaHerring}) are still met,
but the dominant contribution to the integral along $\Sigma_{\rm
chan}$ is awkwardly hidden behind direct-path contributions which are
formally greater but which cancel in a detailed analysis. The best
approach is simply to avoid the whole issue by choosing $\Sigma_{\rm
chan}$ as shown.

As with splittings, the natural next step is to expand $K(y,y',E)$
quadratically about the maximum of its real component. This condition
is given by a periodic orbit of imaginary period (with real
extensions).  There are obvious analogs to the matrices $W$ and $M$
defining the quadratic phase expansions in the double well case, and
these share the symmetry properties derived from $W^*=W^{-1}$.
Finally, we can transform to a phase space representation using the
partial Wigner functions as in section \ref{apsfas}.

To present the final result, we note that the resonance-width-weighted
density of states
\begin{equation}\label{deffGamma}
               f(E) = \sum_n \Gamma_n \delta(E-E_n).
\end{equation}
is approximated in the mean by the term
\begin{equation}\label{truncGamma}
            f_0(E)  =  \frac{1}{2\pi}
                       {e^{-K_0/\hbar}\over \sqrt{(-1)^{\dim-1}
                       \mbox{det}(W_0-I)}},
\end{equation}
defined by the bounce orbit, in analogy with Eq.~(\ref{truncGamma}). 
Then the rescaled resonance width
\begin{equation}\label{defyGamma}
               y_n = \frac{\rho_0(E_n)}{f_0(E_n)}\;\Gamma_n,
\end{equation}
has average value unity and is given by a formula formally identical 
to Eq.~(\ref{finally}).

Finally, we note that it should be obvious how to modify the
discussion when different topographies arise. For example, in the case
of multiple channels one would simply sum over the different bounce
orbits and let the optimal one dominate.  (The same applies to the
case of symmetric double wells when there are multiple tunnelling
routes connecting the wells.)

\section{General Structure of the Result}
\label{Transop}
\label{positivity}

This section consists of two parts. In the first a formal 
interpretation using transfer operators is developed for the 
result, and in the second positivity of the splittings is related to
hyperbolicity of the symplectic matrix $W$.

\subsection{A formal interpretation using transfer operators} 

\noindent
The matrix element we have derived for splittings and resonance widths
can be formulated very concisely in terms of the Bogomolny transfer 
operator and the structures surrounding it \cite{bogomolnytrans}.
This is defined in terms of the real dynamics within a given well,
so let us restrict ourselves to that case initially. For convenience 
of presentation we will use the notation developed for the case of 
splittings in symmetric wells --- subsequent adaptation to resonance 
widths in metastable wells will be obvious. Therefore we start with 
the oriented Poincar\'e section $\Sigl_R$ and denote the usual first 
return map by $\Trl:\,\Sigl_R\to\Sigl_R$. One supposes now that there 
is a Hilbert space $\H(\Sigl_R)$ quantising $\Sigl_R$ and then a 
unitary transfer operator $T:\H(\Sigl_R)\to\H(\Sigl_R)$ is constructed 
within semiclassical approximation whose classical limit is the 
symplectic map $\Trl:\,\Sigl_R\to\Sigl_R$. Matrix elements of $T$ are 
given by a formula like Eq.~(\ref{RLGreen}) except that the real orbits 
and actions defined by $\Trl$ are used. Wavefunctions in $\H(\Sigl_R)$ 
are related to proper wavefunctions by a two-step process: first the 
part of the wavefunction corresponding to traversal of the section in 
the correct sense is extracted, and then the amplitude is modified by 
the square root of the transverse velocity. This corresponds neatly to 
the construction of the arrowed wavefunctions in Eq~(\ref{defE}).  In
particular, we will interpret a directed wavefunction ($\psil_R\!(y)$ 
for splittings and the obvious analog for resonance widths) as the 
wavefunction of a particular element $\ket{n}$ of $\H(\Sigl_R)$. Finally, 
in the transfer operator formalism proper eigenfunctions are related to 
the existence of eigenfunctions of $T$ with unit eigenvalue --- thus we
expect $T\ket{n}=\ket{n}$.  Note that a precise quantum-mechanical
definition of these objects has not yet been given in the completely
general case. This need not concern us, however, because we are using
them for the purposes of interpretation only and ultimately have
well-defined formulas such as Eq.~(\ref{nicellHerring}) to fall back
on.

So far the discussion has been without reference to tunnelling.  We
assume these structures have been calculated in standard semiclassical
approximation within one well, ignoring exponentially small
effects. These are now incorporated using the tunnelling Green's
function construction of the preceding sections.  The Green's function
$\Gtil(y,y',E)$ will be interpreted as the semiclassical quantisation
of a complex mapping $\Tim:\,\Sigl_R\to\Sigl_R$ which is constructed
by using orbits of type $\gamma$ to map $\Sigl_R$ into itself. To
understand the distinction between this map and the usual real
first-return map $\Trl:\,\Sigl_R\to\Sigl_R$, it is useful to think of
a one-dimensional double well. Up to symmetry operations, this has two
periods in the complex time-plane; a real one corresponding to
oscillation within a well and an imaginary one under which any initial
condition is taken to its symmetric partner on the other side of the
barrier (with complex intermediate coordinates even if the initial and
final conditions are real). The maps $\Trl$ and $\Tim$ can then be
interpreted as deformations of these periodic motions into
multidimensional dynamics. It is interesting to compare their
behaviours under conjugation,
\begin{eqnarray}\label{conjinvert}
  	 \Trl^* &=&  \Trl \nonumber \\
	\Tim^* &=& \Tim^{-1},
\end{eqnarray}
which is what one would expect of real and imaginary time evolutions
respectively. Note, however, that $\Trl$ and $\Tim$ are not equal-time
mappings. The bounce orbit $\gamma_0$ defines a real fixed point of
$\Tim$. However, $\Tim$ is in general a completely complex map ---
other initial conditions, even if real, are mapped to points with
complex coordinates.  Thus when we write $\Tim:\,\Sigl_R\to\Sigl_R$,
we must interpret $\Sigl_R$ as being a complex manifold. In principle
we should expect to be able to extend the definition of $\Tim$ for
initial conditions arbitrarily far from $\gamma_0$ by analytic
continuation on $\Sigl_R$, though there will presumably be a rather
complicated branch-point structure in general. In practice
semiclassical approximation requires that we consider $\Tim$ only in
the immediate neighbourhood of $\gamma_0$, which we assume to be free
of branch points and singularities.

Just as there is an operator $T$ quantising $\Trl$, we can construct
an operator $\T$ quantising $\Tim$ --- its matrix elements are given
by the modified Green's function $\Gtil(y,y',E)$ in
Eq.~(\ref{RLGreen}). It is not unitary like $T$ but Hermitean.  In
analogy with Eq.~(\ref{conjinvert}) we compare their properties of
inversion as follows,
\begin{eqnarray}\label{Qconjinvert}
  	 T^\ddagger &=&  T \nonumber \\
	\T^\ddagger &=& \T^{-1}.
\end{eqnarray}
where for any operator we denote with a $\ddagger$ the inverse of its
Hermitean conjugate (to see why we choose this as the quantum analog
of complex conjugation, consider the operator $\exp[-izH]$ for complex
$z$ and Hermitean $H$ and define complex conjugation to be that
operation which sends $z\to z^*$).

We can now formally write the rescaled splitting as,
\begin{equation}\label{absfinally}
	y_n = \frac{\braopket{n}{\,\T\,}{n}}{\Tr\,\T},
\end{equation}
noting that the Weyl symbol of $\T$ is given in Eq.~(\ref{midpoint})
or, in quadratic phase approximation, by Eq.~(\ref{gaussM}). It
should be noted that the state $\ket{n}$ is not normalised
conventionally. Formally, it can be defined to be the state for which
the Wigner function, defined conventionally according to
Eq.~(\ref{defWeyl}) with $\dim$ replaced by $\dim-1$ and coordinates
restricted to the surface of section degrees of freedom, is,
\begin{equation}
	\tilde{\Wt}_n(\zetab) = \rho_0(E_n) \Wl_n(\zetab).
\end{equation}
In practice, calculation of $\ket{n}$ boils down to Eq.~(\ref{defE}),
or some version semiclassically equivalent to it, along with
appropriate redefinition of the normalisation.  Eq.~(\ref{MCPW})
indicates that $\ket{n}$ is normalised so that the average value of
this Wigner function is unity or, more generally, such that its
coefficients in a generic basis for $\H(\Sigl_R)$ average to one.

While Eq.~(\ref{absfinally}) can be used as it stands, it is
considerably easier to implement if a Gaussian approximation is used,
as in Eqs.~(\ref{GHerring}) and (\ref{finally}). This amounts to
substituting for $\T$ the quantisation of the linear symplectic map
$W$ instead of that of the full nonlinear map $\Tim$. Before writing
this explicitly, it is useful to be more specific about the structure
of $W$.  Define $\Orient$ to be the symplectic transformation which
inverts an eigenvector of $W$ if the corresponding eigenvalue is
negative and leaves it invariant otherwise.  Then $\Orient$ commutes
with $W$ and $\Orient W$ is hyperbolic, {\it i.e.}, has only positive
real eigenvalues. We then implicitly define a real positive definite
symmetric matrix $B$ through
\begin{equation}\label{sepW}
	W = \Orient e^{-i JB},
\end{equation}
That $B$ defined in this way is symmetric follows from the fact that
$W$ is symplectic and that it is real follows from
Eq.~(\ref{revcon}). Positivity comes from more specific dynamical
properties of the system and is discussed further in the next
subsection. In other words, we consider $W$ to be generated by
evolution under the positive quadratic Hamiltonian,
\begin{equation}
	h(\zetab) = \ha \zetab^T B \zetab .
\end{equation}
for the imaginary time $\Delta t=-i$, followed by a reorientation 
$\Orient$ of certain eigenvectors to allow for negative eigenvalues.

The Gaussian approximation now corresponds simply to the substitution,
\begin{equation}\label{gaussT}
	\T' =  \U \exp[-\hh/\hbar].
\end{equation}
for $\T$.  In the exponent, $\hh$ is a positive-definite hermitean
operator which is quadratric in the operators $\hat{y}$ and
$\hat{p}_y$ and whose Weyl symbol is $h(\zetab)$. We will write $\hh=
\hat{\zetab}^T B \hat{\zetab}/2$ where $\hat{\zetab}$ is a vector of
operators defined in obvious analogy with Eq.~(\ref{spinor}).  $\U$ is
a unitary matrix quantising $R$ --- its sign is fixed by demanding
that its Weyl symbol be positive. Up to normalisation, $g_M(\zetab)$
is the Weyl symbol of this operator, as can be checked explicitly.

We note finally that in two degrees of freedom the spectrum of $\T'$
is the geometrically decaying sequence
\begin{equation}\label{spectrum}
	\mbox{Spectrum}\,[\,\T']
	 = \left\{ (\pm 1)^k e^{-(k+1/2)\hbar\alpha/\hbar}
		 \right\}_{k\ge 0}
	 = \left\{ |\Lambda|^{-\ha}\Lambda^{-k} \right\}_{k\ge 0},
\end{equation}
of powers of the eigenvalue $\Lambda=\pm e^{\alpha}$ of $W$ which is
larger in magnitude, $\alpha=\left[\det B\right]^{1/2}$ being the
frequency associated with real-time evolution under $h(\zetab)$.  In
higher dimensions the spectrum of $\T'$ consists of products of such
powers of eigenvalues.

\subsection{Reflections, inversions and negative splittings}

In certain situations, one expects that the $y_n$'s should
systematically be positive. This is certainly the case for resonance
widths, for example. There is also often an expectation in the case of
symmetric double wells that the even state should be at a lower energy
than the odd state, as is the case in one-dimensional wells and in the
system illustrated in section~\ref{theoryforthemean}.  It is not
always so, however. For example, when the symmetry underlying the
splitting is a complete spatial inversion $\R:\x\to-\x$, negative
splittings are routinely encountered for excited states, as discussed
in appendix~\ref{Invert}.  The following question now arises. When is
the matrix element underlying $y_n$ inheherently positive and how can
we relate this positivity to to structural and dynamical properties of
the problem?

The answer is straightforwardly encoded in Eq.~(\ref{spectrum}) and
can be summarised by the following statement,
\vspace*{3pt}\\
{\bf Assertion:} {\it In Gaussian approximation, $\T'$ is a positive
definite operator whenever $W$ is hyperbolic, that is, whenever $W$
has only positive eigenvalues. In particular, this is expected to be
the case for resonance widths and for splittings in symmetric double
wells when the symmetry is reflection in one cartesian coordinate.
When $W$ has negative eigenvalues, so does $\T'$, and we expect both
positive and negative values for the $y_n$'s. This is the case, for
example, in a symmetric double well whose symmetry is a complete
inversion.}
\vspace*{3pt}\\
Hyperbolicity of $W$ is to be expected for a complex orbit crossing a
barrier in a potential and computed with respect to a rigidly
translated basis (see below). When nonhyperbolicity arises, it is
because the coordinate frame with respect to which the final
displacement is measured has undergone a reorientation with respect to
the rigidly translated basis. In the case of symmetric double wells,
such a frame transformation would be induced by the application of
$\sigma$ to map the coordinate system of one section onto its
symmetric partner as in the discussion preceeding Eq.~(\ref{defG}). In
the case of metastable wells, there is no such transformation and $W$
is always hyperbolic.

To understand why we expect $W$ to be hyperbolic in the absence of
frame rotation, consider the special case of a potential barrier with
a symmetry axis, so that $\gamma_0$ is quasi-one-dimensional, as in
Fig.~\ref{potview}. Then $W$ is the solution of the differential
equation,
\begin{equation}\label{intW}
	\frac{\d W}{\d\tau} = -iJ [H'']_\perp W,
\end{equation}
where $[H'']_\perp$ is the Hessian matrix of the Hamiltonian in
cordinates transverse to $\gamma_0$. We expect the potential, and
therefore $[H'']_\perp$, to be positive definite underneath the
barrier and it then seems clear that the solution of Eq.~(\ref{intW})
would be of the form $e^{-i JB}$, with positive real symmetric $B$, as
in the previous subsection.  We claim without proof that this is the
case and that it also holds in the more general case when $\gamma_0$
is not on a symmetry axis.

It should be acknowledged that, while positivity of $\T'$ is certainly
a necessary condition for the positivity of the $y_n$'s it cannot yet
be taken as sufficient. It will be seen in explicit implementation in
the next section that it is often the case that there is significant
cancellation in the matrix element, leaving a very small result, and
corrections could in principle tip the balance and result in negative
splittings. We do not believe this to be the case in practice, but
have not rigourously proved otherwise.

It is interesting to note how positivity arises in the Wigner-Weyl
formalism of section~\ref{apsfas} --- irrespective of the
hyperbolicity of $W$, the splitting is given by the overlap of a
positive Gaussian $g_M(\zetab)$ with the Poincar\'e-Wigner function,
and at first sight it is not obvious why and when this overlap would
be negative.  In broad terms, the solution is as follows. If
$g_M(\zetab)$ is sufficiently narrow (or equivalently if $M$ is
sufficiently large) then one can see that it might be supported in a
region where the Wigner function is negative and a negative overlap
might emerge. On the other hand, if $g_M(\zetab)$ is very broad (so
$M$ is very small), we might expect the overlap to wash out any
negative regions and only positive overlaps to emerge. Unsurprisingly,
the smallness or largeness of $M$ is linked to the hyperbolicity of
$W$.

To simplify the discussion, let us restrict ourselves 
to $d=2$. We can then give a very precise criterion for $M$ in 
order that the overlap be positive. This is summarised by the 
following statement,
\vspace*{3pt}\\
{\bf Assertion:} {\it The overlap of $g_M(\zetab)$ with an arbitrary
Wigner function is always positive precisely when $\det M <1$, in
which case $g_M(\zetab)$ is the Weyl symbol of a positive definite
operator of the form $\exp[-\hat{\zetab}^T B\hat{\zetab}/2\hbar]$. In
the limiting case $\det M=1$, $g_M(\zetab)$ is the Weyl symbol of a
pure state and, when $\det M >1$, it is the Weyl symbol of an operator
of the form $\U \exp[-\hat{\zetab}^T B\hat{\zetab}/2\hbar]$ where $\U$
is a unitary operator quantising the inversion $R=-I$.}
\vspace*{3pt}\\
The first part of this statement comes simply from calculating the
Weyl symbol of $\exp[-\hat{\zetab}^T B\hat{\zetab}/2\hbar]$, which is
proportional to,
\begin{equation}
	\exp\left[ -\tanh(\alpha/2) 
               \frac{\zetab^T B\zetab}{\hbar\alpha}\right],
\end{equation}
where $\alpha^2=\det B$, and equating its exponent with that of
$g_M(\zetab)$. A solution can be found precisely when $\det M<1$ (note
that $\det[\tanh(\alpha/2)B/\alpha] = \tanh^2(\alpha/2)<1$). The
marginal case $\det M=1$ is not directly relevent to tunnelling
calculations, but was included for completeness --- it would
correspond to the limit $\Lambda\to\infty$. The case $\det M>1$ is
easily seen to correspond to the insertion of an inverse hyperbolic
matrix $W$ in Eq.~\ref{Mdef}. Note that replacing $W$ with $-W$ in
that equation leads to the following replacement for $M$,
\begin{equation}
	W\longrightarrow W'=-W \qquad\Rightarrow\qquad 
                 M\longrightarrow M'=J^T M^{-1} J.
\end{equation}
In particular, $\det M' =1/\det M$, so $\det M>1$ corresponds to an
inverse hyperbolic matrix as claimed. It can also be verified by
explicit calculation that if $g_M(\zetab)$ is the Weyl symbol of
$\exp[-\hat{\zetab}^T B\hat{\zetab}/2\hbar]$, then $g_{M'}(\zetab)$ is
the Weyl symbol of $\U\exp[-\hat{\zetab}^T B\hat{\zetab}/2\hbar]$.  In
other words, $g_{M'}(\zetab)$ is the Moyal product of $g_M(\zetab)$
with the Weyl symbol of $\U$.

Finally, we note that in the case $d=2$, there are two possible
symmetries for a double well. Reflection through an axis defines a
hyperbolic $W$ and positive splittings, whereas inversion through a
point defines an inverse hyperbolic $W$ and produces a mixture of
positive and negative splittings (appendix~\ref{Invert}).

\section{An Explicit Implementation} 
\label{thesystem}

\noindent
To test the results, we will work with the potential defined in
Eq.~(\ref{fop}) of section~\ref{theoryforthemean}, and illustrated in
Fig.~\ref{potview}.  Classical motion is constrained to one or other
of the wells for energies in the range $0<E<1$ and we expect nearly
degenerate doublets in the quantum spectrum.  This is a convenient
choice for the numerical calculation of quantum spectra since the
potential is polynomial in $x$ and $y$ and the Hamiltonian is then
banded in a basis of harmonic oscillator eigenstates, allowing large
bases and therefore accurate diagonalisation. In addition, the
classical dynamics is largely chaotic over a wide range of energies.

In a previous publication \cite{us}, we discussed tunnelling rates in
this potential for the case $\mu=0$. This choice has the disadvantage
that the potential does not have a generic behaviour at the top of the
barrier; instead of a saddle point, one finds a degenerate ridge
seperating the two wells. This leads to nongeneric behaviour of
tunnelling rates for states just below this critical energy. A
corrective analysis is certainly possible but would be specific to
that case and not particularly interesting in its own right. We
therefore prefer to avoid this situation by choosing $\mu>0$.  On the
other hand, large values of $\mu$ induce large regular regions in
phase space and so we limit ourselves to values of $\mu$ that are
relatively small, specifically $\mu=1/10$.

\begin{figure}[h]
\vspace*{0.5cm}\hskip .5cm 
\psfig{figure=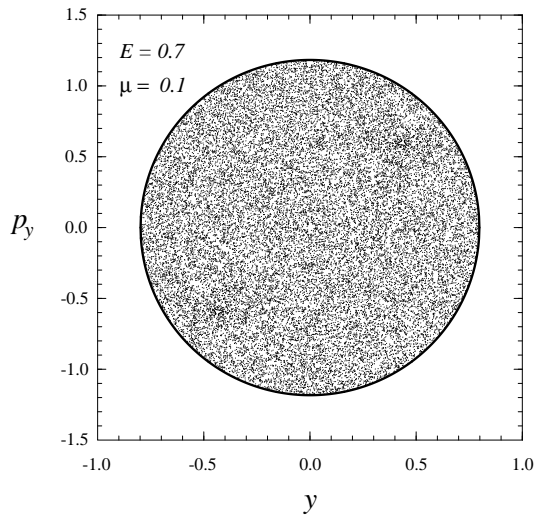,height=2.5in}
\vspace*{0.4cm}
\caption
{\small Classical dynamics is illustrated in the form of the $x=1$, $p_x<0$
Poincar\'e section for $E=0.7$ and $\mu=0.1$, which is the case used
for numerical calculation. The section is almost completely chaotic,
with no visible islands.}
\label{poinc}
\end{figure}

The classical motion is then predominantly chaotic as long as the
energy is not too small (typically, $E>0.3$).  A typical Poincar\'e
plot is shown in Fig.~\ref{poinc}, for $E=0.7$. The section is defined
by plotting $y$ and $p_y$ whenever $x=1$ and $p_x<0$. The real
periodic orbit which connects with the bounce orbit intersects this
Poincar\'e section at the origin.  Although some small regular islands
are present, they are hardly visible and the dynamics is dominantly
chaotic. It is an important feature because we can then expect the
wavefunctions to be ergodically spread throughout the space so that
they can ``access'' the central tunnelling route $\gamma$ discussed
above. For nonchaotic systems, wavefunctions are typically strongly
localised in phase space and the previous analysis would be relevant
only to a subset of the states, assuming the real extension to the
bounce orbit to be unstable.  If it is stable, the region around it
supports eigenstates whose energies and splittings can be extracted
semiclassically \cite{us} without any reference to the wavefunction at
all. The presence of chaos will also be important to justify various
arguments necessary in a statistical analysis of the splittings
\cite{us''}.

In order both to invoke this formalism and to test it, we need to
diagonalise the quantum problem numerically. On the quantum-mechanical
level, there is an additional parameter which appears, namely Planck's
constant. The classical mechanics is independent of $\hbar$
while it does depend on the choice of energy. For this reason, the
analysis is much cleaner if we hold the energy fixed at $E$ and find
those values of $\hbar$ for which it is an eigenvalue  --- {\it
i.e.} perform a numerical quantisation of $\hbar$. This can be done by
considering Schr{\"o}dinger's equation to be a generalised eigenvalue
equation for $\hbar$. Defining the momentum squared operator
$\hat{p}^2/2=\hbar^2\hat{T}$, we have
\begin{equation} \label{gee}
\opket{(E-\hat{V})}{\psi_n} = \hbar_n^2\opket{\hat{T}}{\psi_n}
\end{equation}
This is readily solved in a two-dimensional harmonic oscillator basis
using standard numerical routines which make use of the fact that
$\hat{T}$ is a positive-definite operator. In what follows we will
analyse the spectrum of the reciprocal quantity 
\begin{equation}
                           q=\frac{1}{\hbar}. 
\end{equation}
The even and odd spectra are very close but there are small tunnelling
splittings $\Delta q_n$ between the respective states, just as for the
energies.

The only remaining issue is that the theory developed was for
splittings in energy so we would like to convert our $q$-splittings
into energy splittings. This can be done using first order
perturbation theory.  Each doublet consists of one state which is even
with respect to reflections in $x$ and another which is odd; the even
state always being observed to have the lower energy.  The same
property holds as a function of $q$; the even states have slightly
smaller values than the odd states. This can be understood if we think
of each state corresponding to a parametric curve in $E-q$ space. The
doublets are then pairs of very closely spaced curves in this
space. Because $\hat{T}$ is a positive-definite operator, a first
order perturbation calculation shows that $\dydxh{E}{q}<0$. From this
it follows that if the even state of each doublet has a smaller value
of energy at fixed $q$, then it also has a smaller value of $q$ at
fixed energy. First order perturbation theory then gives
\begin{equation}\label{pertth}
\Delta E_n = {2\over q_n^3}\braopket{\psi_n}{ {\hat T}}{\psi_n}
\Delta q_n.
\end{equation}
The matrix element is readily calculated, since the wavefunction
$\psi_n$ is known and so, given the $q$ splitting, we can readily
determine the energy splitting. Any approximation for $\psi_n$ that is
more accurate than semiclassical error is acceptable. In practice this
would mean performing a numerical calculation with a basis confined to
one well.  The relation (\ref{pertth}) is valid only to leading order
in $\Delta q_n$, however this level of approximation is consistent
with earlier arguments.

\begin{figure}[h]
\vspace*{0.5cm}\hskip 0.2cm 
\psfig{figure=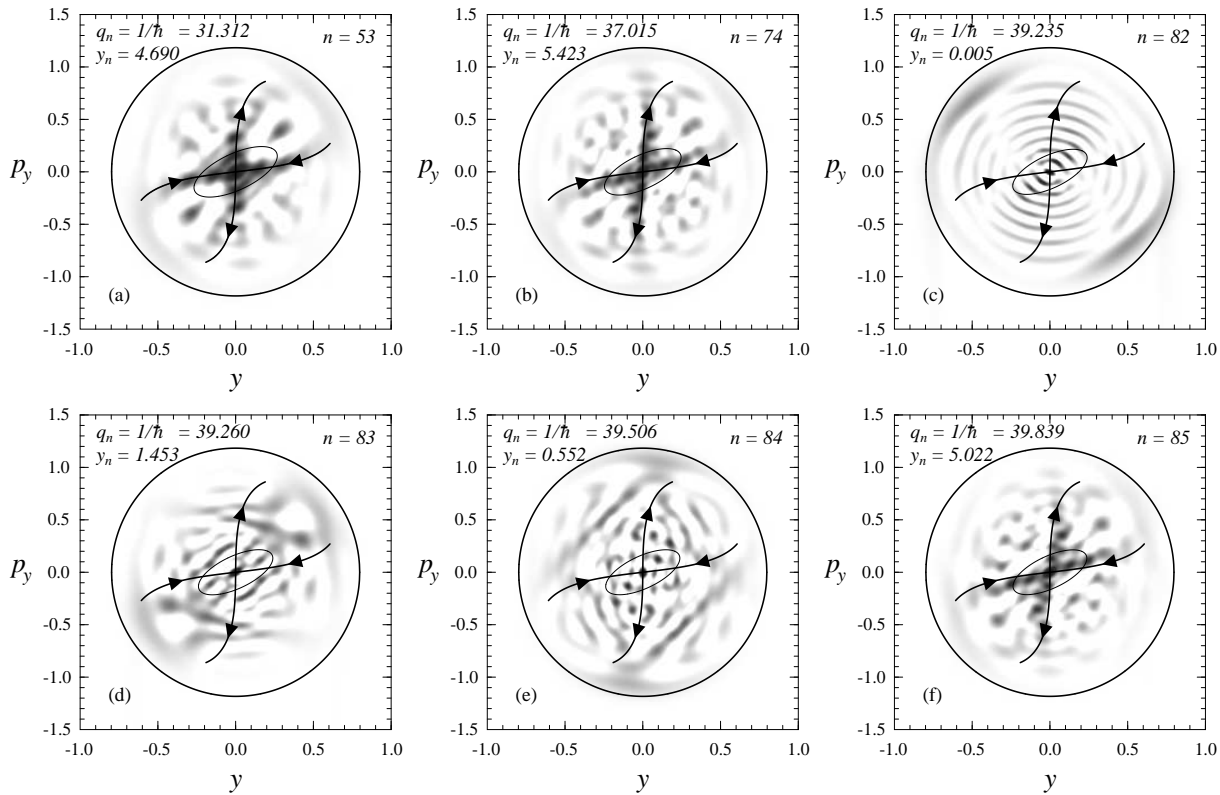,height=3.5in}
\vspace*{1.5cm}
\caption
{\small Some explicit examples of Poincar\'e-Wigner functions,
obtained by quantising $\hbar$ for the parameter values illustrated in
the classical surface of section of Fig.~\ref{poinc} (negative values
are represented by white space). Cases (a), (b)
and (f) have exceptionally large splittings and correspond to states
scarred by the periodic orbit at the center. Their Wigner functions
are concentrated along its stable and unstable manifolds, shown as the
heavy curves emerging from the origin. State (c) has an exceptionally
small splitting, and (d) and (e) are ``typical''. Also shown in each
case is the ellipse along which the Gaussian $g_M(\zetab)$ has fallen
by a factor $e^{-1/2}$ --- its area is $\pi\hbar/2\left[\det M\right]^\ha
\approx 5\hbar$.  This elliptic structure is defined by
the complex monodromy matrix $W$ and is distinct from the hyperbolic
structure of the invariant manifolds associated with the real matrix
$S$ which linearises dynamics about the center. This structural
dichotomy emerges even though the eigenvalues of both matrices are on
the real axis.}
\label{wigner}
\end{figure}

We have tested the formalism in describing the splittings for the
choice $E=0.7$, $\mu=0.1$ and even $y$-parity.  We have also worked
predominantly with the surface of section defined by $p_x<0$ and
$x=1$, though some results will also be presented for the choice
$x=1/2$ (these sections are respectively referred to as $\Sigma_a$ and
$\Sigma_b$).  For the former choice we have that the real time of
evolution for the orbit $\gamma_0$ is $t_0=0.93$ while the imaginary
time ({\it i.e.} for the bounce orbit) is $-i\tau_0=-1.23i$.
Therefore, we start at $\Sigma_R=\Sigma_a$ with $y=p_y=0$, $x=1$ and
$p_x$ given by energy conservation (with $p_x<0$). We integrate these
initial conditions for a time $t_0$, which brings us to the barrier.
Following this we integrate for a time $-i\tau_0$, which carries us
across to the left-hand well.  Finally, we integrate for a time
$-t_0$, bringing us to $\Sigma_L$, with the final velocity pointing
once again towards the barrier. The resulting point is at $y=p_y=0$
and $x=-1$ and $p_x>0$, which is the mirror image in phase space of
the initial point.  While doing these integrations, we also keep track
of the complex monodromy matrix $W$.  We then construct the matrix $M$
as given by Eq.~(\ref{Mdef}) and thereby construct the Gaussian
function on the Poincar\'e surface defined in Eq.~(\ref{gaussM}).

As a final step, we evaluate the partial Wigner function
Eq.~(\ref{halfassedWigner}) of the wavefunction $\psi_n$. This can be
done efficiently by expressing the state in a position representation,
using the expression (\ref{defE}) to determine the semiclassical
wavefunction defined on the section and then performing a fast fourier
transform to get the phase space representation. Since the Weyl
transform of the Green function above was calculated using the central
Gaussian approximation, we do the analogous thing for the wavefunction
by taking the velocity factors appearing in (\ref{halfassedWigner}) to
be independent of $y$ and $y'$ and simply equal to the values for the
central bounce orbit. A few cases are shown in Fig.~\ref{wigner} where
it is shown that states with relatively large splittings are strongly
scarred while the other states are not. In fact, for the most strongly
scarred states, one can even observe amplitude extending out along the
stable and unstable manifolds of the orbit. This supports the
statement that the quantity $y_n$ and the phase space integral leading
to it represent a quantification of the extent of scarring.

We then integrate this function with the Gaussian function as
prescribed in Eq.~(\ref{finally}). One final complication is that the
symmetry in $y$ necessitates some additional analysis. In particular,
states even in $y$ typically have larger splittings than states odd in
$y$ simply because they have more amplitude on the tunnelling path.
This property is handled automatically in the first equation of
Eq.~(\ref{finally}) since the factor $\Wl_n\!(\zetab)$ will typically
be larger for the even states than for the odd. Therefore in
evaluating the raw splittings $\Delta E_n$, we use the same prefactor
$f_0$ for both symmetry classes; the one described in section
\ref{theoryforthemean}. However, we will want to normalise by
different amounts for the two classes to get $y_n$ values which
separately average to unity and so we need to use parity-specific
forms for $f_0$ and $\rho_0$. Specifically, we use the relation
$y_n=\rho^\pm_0(E_n)\Delta E_n/f^\pm_0(E_n)$ to get the normalised
splittings, where the $\pm$ refers to the $y$ parity and the means for
determining these functions is given in
appendix~\ref{symmetrydecomposition}. In this way, we finish with
normalised splittings which average to unity for both symmetry classes
even though the average splittings before normalisation are different.

Again, we stress that this only makes use of the wavefunction in the
classically allowed region and is not a particularly demanding
numerical task.

\begin{figure}[ht]
\vspace*{0.5cm}\hskip 2.5cm 
\psfig{figure=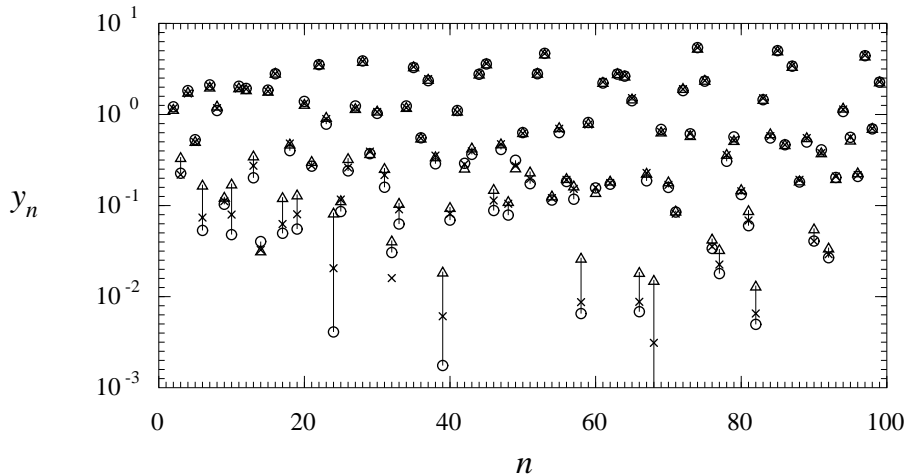,height=2.5in}
\vspace*{0.5cm}
\caption{\small Some explicit numerical results, calculated in the 
Poincar\'e section $\Sigma_R=\Sigma_a$ defined by $x=1$. The
horizontal axis is just an index labelling the states and the vertical
axis shows the rescaled splittings on a logarithmic scale. The exact
quantum results are shown as circles. There are two separate 
semiclassical calculations, represented by triangles and crosses.
The triangles represent the simplest calculation using the 
Gaussian approximation --- these are joined by lines
to the quantum results so as to guide the eye. Agreement
is generally good, except for very small splittings, which tend to be
overestimated by the approximation, though the situation improves
somewhat with increasing quantum number. These smaller splittings 
are often better estimated by the second semiclassical approximation,
represented by the crosses, in which the exact action and amplitude 
of the complex orbit is used. This is the noncentral analysis 
referred to in the text.}
\label{showyourstuff}
\end{figure}

In Fig.~\ref{showyourstuff} we show the results of this procedure
(triangles), for the first 100 states even in $y$, and make a
comparison to results of a completely quantum-mechanical
diagonalisation (circles).  The agreement is rather impressive,
especially for the large splittings.  For the bulk of the states,
those with rescaled splittings of order unity or greater, the
agreement is typically within a few percent.  It should be remarked
that the agreement does improve systematically as one climbs higher in
the spectrum.

The states with small splittings are anti-scarred --- having
suppressed amplitude along the central periodic orbit. This is
problematic since our Gaussian approximations assume that it is the
central region which dominates the integral. While certainly true for
the bulk of the states, it may not be true for the handful of
anti-scarred ones. To test this, we went beyond the Gaussian
approximation. Specifically, we numerically found the complex orbit
connecting each pair of points $(x',y')$ and $(x,y)$ in the two
wells. Each such orbit has an action, a stability matrix and initial
and final $x$ velocity. We fed this information directly into
Eqs.~(\ref{defE}) and (\ref{RLGreen}). From those, we could then
determine the corresponding Weyl and Wigner functions using
Eqs.~(\ref{halfassedWeyl}) and (\ref{halfassedWigner}). These are
still real functions due to the hermiticity properties satisfied by
$\E$ and $\Gtil$. We then used Eq.~(\ref{WHerring}) to determine the
splittings. The results of this analysis are represented by the
crosses in Fig.~\ref{showyourstuff}. The agreement is dramatically
improved for those states with small $y$ values --- as expected. It is
also generally improved throughout the spectrum, except for the very
first few states. It should be stressed that these are still only
approximate since there are still semiclassical approximations being
made in the analysis. Note that, this procedure is far more difficult
than just using the central approximation since thousands of complex
orbits must be found rather than just the one trivial central orbit. A
compromise which seemed to improve the results significantly while
still requiring relatively little work was to expand $\dot{x}$ and
$\dot{x}'$ as quadratic forms in $y$ and $y'$ about the central orbit
and using these expressions in Eq.(\ref{defE}).  However, for the sake
of brevity we do not include these numbers. It should be mentioned
that in the semiclassical limit $\hbar\rightarrow 0$, we expect that
only a vanishing fraction of the states will be so anti-scarred as to
require this non-central analysis.

In Table~\ref{tableyourstuff}, we give a more detailed comparison of
results for some selected states --- the first five and a second
sequence of five higher in the spectrum. A comparison is made between
quantum calculation and of our matrix element evaluated in two
distinct sections, $\Sigma_a$ and $\Sigma_b$ as well as with using the
non-central analysis discussed in the preceding paragraph (for the
section $\Sigma_a$). Note that the results from $\Sigma_a$ and
$\Sigma_b$ are different, but consistent within the level of
approximation achieved while those from the non-central analysis
improve significantly the small y values. On the whole, results from
$\Sigma_a$ or $\Sigma_b$ seem more or less equivalent. However, we
have noticed that the smallest splittings tend to be better reproduced
by the $\Sigma_b$-calculation.

A possible explanation is that semiclassical errors increase with
increasing real time propagation at either end of the barrier
crossing.  This suggests that it might be advantageous to perform our
Gaussian overlap of the wavefunction as close as possible to the
turning point.  As we have discussed, the derivation we have outlined
is problematic in this limit. However, we expect that the final
result, being formally canonically invariant, should be equally valid
as long as a consistent generalisation of the Poincar\'e-Wigner
function is employed. If calculation is then done in a Poincar\'e
section that remains well-defined at the turning point, we expect the
main formula to remain valid. The main ingredient in such a procedure
is to perform a symplectic rotation of the quantities entering main
result --- since the Wigner-Weyl calculus is covariant under
symplectic rotation this is certainly possible, though we have not
implemented it in detail.

We have verified that the formalism works for other choices of energy,
parameter values and $y$ parity, although, for sake of space, we do
not present the corresponding results.

\begin{table}[ht]
\begin{center}
\begin{tabular}{|r|r|r|r|r|r|}
\hline
$n$ & $q_n$ \hfill & $y_n$ [QM]  
& $y_n$ [$\Sigma_a$] & $y_n$ [$\Sigma_b$] & $y_n$ [NC]\\
\hline
\hline
   1 &  2.361  &  0.5475  &  0.4628  &  0.3801  &  0.4408\\
   2 &  5.067  &  1.2132  &  1.1521  &  1.1238  &  1.1276\\
   3 &  5.924  &  0.2267  &  0.3407  &  0.2315  &  0.1437\\
   4 &  7.896  &  1.8310  &  1.7773  &  1.8864  &  1.6533\\
   5 &  8.442  &  0.5282  &  0.5122  &  0.5940  &  0.4227\\
  81 & 38.795  &  0.0601  &  0.0891  &  0.0742  &  0.0653\\
  82 & 39.235  &  0.0050  &  0.0132  &  0.0055  &  0.0048\\
  83 & 39.260  &  1.4533  &  1.4842  &  1.6126  &  1.4004\\
  84 & 39.506  &  0.5517  &  0.6219  &  0.5964  &  0.5484\\
  85 & 39.839  &  5.0215  &  5.0546  &  5.1767  &  4.9092\\
\hline
\end{tabular}
\end{center}
\vspace*{0.5cm}
\caption{\small A detailed comparison of rescaled splittings for the
first five doublets, and for another sequence of five higher in the
spectrum. The third column gives results of an exact quantum
mechanical diagonalisation. Semiclassical approximations were
constructed using Poincar\'e sections at $x=1$ ($\Sigma_a$) and
$x=1/2$ ($\Sigma_b$), results of which are respectively shown in the
fourth and fifth columns. The final column shows the results using
$\Sigma_a$ but with the non-central analysis. Clearly there is no
formal reason to expect agreement for the lowest states, and the
comparison of the first five should be regarded as a case of
experimentation rather than verification.}
\label{tableyourstuff}
\end{table}

Finally, we stress that while a very precise quantum mechanical
calculation has been possible for the potential in Eq.~(\ref{fop}),
this in large part due to its polynomial nature and is not at all a
straightforward task in general. In other potentials, for which
equivalently large bases are not feasible, it may well be impossible
to calculate energy levels to the precision necessary to determine
level splittings when these are very small.  On the contrary,
implementation of the matrix element formula will hardly be different
in terms of difficulty. It is for such systems that the formula will
be of greatest help as a numerical method.

Although we have not yet tested the formalism for the case of
resonances in a metastable well, this is something we intend to do in
the near future.

\section{Conclusion}
\label{conclusion}

\noindent We have shown that an energy-level splitting or a resonance 
width can be effectively calculated as a matrix element of an
explicitly-constructed tunnelling operator, defined on the
quantisation of a Poincar\'e section and having a Gaussian
kernel. This tunnelling operator is completely determined by the
action of an imaginary-time ``bounce orbit'' and a linearisation of
dynamics around it. Besides providing a relatively painless numerical
algorithm for calculating splittings (otherwise a delicate task
numerically), we hope that the formal simplicity of the result will be
useful in theoretical analysis, especially statistical
\cite{us''}.

Even though the bounce orbit crossing the barrier is complex, it
defines a completely real orbit in each well, sharing a turning point
with each of them in the case of time-reversal symmetry. The matrix
element measures the weight of the wavefunction about these real
extensions --- the complex nature of the crossing enters only in the
definition of the Gaussian kernel, which is constructed using the
complex monodromy matrix $W$ of the bounce orbit.  Though complex, $W$
defines a real positive-definite quadratic ``Hamiltonian'' $h(y,p_y)$
defined on a Poincar\'e section, associated with which is an elliptic
structure centered on the real orbit.  The behaviour of a
Poincar\'e-section Wigner function, within an area of $O(\hbar^{d-1})$
defined by these elliptic domains, suffices to determine the
splittings (except for the strongly anti-scarred ones for which a
non-central analysis seems necessary).  In particular, in the common
case where the real extension is periodic, we have a simple
interpretation of the tunnelling rate as a scarring weight. We stress
that computation of the classical data entering in this formula is of
the same degree of simplicity for all systems or parameter values,
while a completely quantum calculation of the splitting can easily
become impractical if the splitting is extremely small or efficient
diagonalisation methods are unavailable.

\vspace{1cm}

\noindent {\bf Acknowledgements}\\
\noindent
We would like to thank Eugene Bogomolny, Oriol Bohigas, Dominique
Delande, David Goodings, Jim Morehead, Steve Tomsovic, Andr\'e Voros
and Michael Wilkinson for useful discussions. N.D.W.  acknowledges
support from the Natural Sciences and Engineering Research Council of
Canada.

\appendix

\section{Symmetry Decomposition}
\label{symmetrydecomposition}

\noindent
The example we choose to study has an additional reflection symmetry
in the $y$ direction as well as the reflection symmetry in $x$ which
is responsible for the presence of a double well. To evaluate the
quantities $y_n$ we need to normalise by appropriately symmetrised
functions $\rho_0^\pm(E)$ and $f_0^\pm(E_n)$ as discussed in 
section~\ref{thesystem}. In this appendix we explain how this is done.

\subsection{Thomas-Fermi calculation}

\noindent
We will first work with the integrated densities of states which can
then be differentiated with respect to either energy or $q$ to obtain
the corresponding densities.  In fact, we need the densities
corresponding to a given value of the $y$ parity, $\pi_y=\pm 1$. The
problem of symmetry decomposition of the smooth density of states has
been discussed in various papers both for potential systems
\cite{pottf} and billiards \cite{nicholas}. Our case is rather simple
since it is a simple parity group and the result is,
\begin{equation} \label{decomp}
N_0^\pm(E,q) = {1\over 2} \left(N_I (E,q) \pm N_R(E,q)\right),
\end{equation}
where $N_I (E,q)$ and $N_R (E,q)$ correspond to the group elements of
identity and reflection through the $x$ axis. (The reflection operator
through the $y$ axis ({\it i.e.} $\pi_x=\pm 1$) plays no role in this
calculation since the axis lies outside of the classically allowed
domain.)

To leading order in $q$,
\begin{equation} \label{ni}
N_I (E)  = {q^2\over 4\pi^2} \int \d z \; \Theta(E-H(z))
\end{equation}
where $z$ collectively represents the phase space coordinates
$(x,y,p_x,p_y)$. The integral is greatly facilitated by remarking that
the potential (\ref{fop}) is quadratic in $y$. Energy conservation
then gives
\begin{equation} \label{rewrite}
E-U(x) = {p_x^2\over 2} + {p_y^2\over 2} + (\lambda x^2 +\mu) y^2,
\end{equation}
where $U(x)=(x^2-1)^4$.  This is just the equation for an ellipsoid
with semiaxes $(a,a,b)$ where $a=\sqrt{2(E-U(x))}$ and
$b=a/\sqrt{2(\lambda x^2+\mu)}$; its volume is then $4\pi
a^2b/3$. Therefore in (\ref{ni}) we have only to numerically evaluate
the $x$ integral,
\begin{equation} \label{whatsleft}
N_I(E) = {2q^2\over 3\pi}\int_{x_-}^{x_+} \d x 
\sqrt{(E-U(x))^3\over\lambda x^2 + \mu},
\end{equation}
where $x_\pm=\sqrt{1\pm E^{1/4}}$. The reflection operator contributes
an amount \cite{pottf} 
\begin{eqnarray}
N_R (E) & = & {q\over 4\pi}\int \d x \d p_x \Theta(E-(p_x^2/2+U(x)))\nonumber\\
        & = & {q\over 2\pi}\int_{x_-}^{x_+}\d x\sqrt{2(E-U(x))}. \label{ns}
\end{eqnarray}

To determine the densities of states in $E$ or in $q$, we
differentiate (\ref{decomp}) with respect to the appropriate variable.
This simple exercise is left to the reader, resulting in quadrature
which are readily evaluated numerically.  It is clear that this only
applies for energies below the barrier. Also, we have only considered
the right well, obviously there is an identical contribution from the
left well. However, strictly speaking, we are usually interested in
the density of doublets and not the density of states. There are half
as many doublets as states which compensates for the factor of two
from not including the left well.  Finally, these expressions are
asymptotic series in $\hbar^2=1/q^2$ which we have truncated at the
first term. The first correction to $N_I(E)$ is a constant which is
typically small for potentials and furthermore has no effect on the
density of states at finite energy and $q$.

\subsection{Stability factor calculation}

\noindent
The function $f_0(E)$ defined in Eq.~(\ref{trunc}) comes from
considering just the pure bounce orbit. In the presence of a
reflection symmetry in $y$, this orbit lies on the symmetry axis and
its amplitudes decomposes between the even and odd states in a
specific way \cite{bent}. We start by defining the larger eigenvalue
of $W_0$ appearing in Eq.~(\ref{trunc}) by $\Lambda$. Then we can write
the stability factor (with $\dim=2$) as
\begin{eqnarray}
{1 \over \sqrt{-\mbox{det}(W_0-I)}} 
& = & {\Lambda^{1/2} \over \Lambda-1} \nonumber\\
& = & {\Lambda^{3/2} \over \Lambda^2-1} + {\Lambda^{1/2} \over \Lambda^2-1}.
\label{stabfactor}
\end{eqnarray}
The first equation comes trivially from the determinant while the
second comes from expanding out the factor of $1/(\Lambda-1)$ as a
geometric series in $1/\Lambda$, separately collecting the even and
odd powers and then resumming the two series. In any event, it is
trivial to see that the second line equals the first regardless of how
it was derived.

We now follow reference~\cite{bent} and identify the first term as
giving the contribution to the even states and the second term as
giving the contribution to the odd states.  The exponential factor in
Eq.~(\ref{trunc}) remains the same and so we identify
\begin{eqnarray}
f_0^+(E)
& = & {1 \over \pi}{\Lambda^{3/2} \over \Lambda^2-1}e^{-K_0/\hbar} \nonumber\\
f_0^-(E)
& = & {1 \over \pi}{\Lambda^{1/2} \over \Lambda^2-1}e^{-K_0/\hbar}.
\label{symmetrisedf0s}
\end{eqnarray}
We observe that the even splittings are, on average, larger by a
factor of $\Lambda$.

\section{Thomas-Fermi for Poincar\'e-Wigner functions}
\label{TFPW}

\noindent
In this appendix we outline how to derive the microcanonical
background of the Poincar\'e-Wigner functions [Eq.~(\ref{MCPW})] from
that of the standard Wigner functions [Eq.~(\ref{MCW})], which we
assume to be known \cite{gutz}.

The main ingredient is the following transformation rule between
ordinary Wigner functions and partial Wigner functions,
\begin{equation}
              \Wt_\psi(x,x',y,p_y) = (2\pi\hbar)^\dim \int \d p_x\; 
		       e^{-ip_x(x-x')/\hbar}\; 
		W_\psi\left((x+x')/2,y,p_x,p_y\right),
\end{equation}
straightforwardly deduced from their definitions in Eqs.~(\ref{Weylmat})
and (\ref{halfassedWeyl}). In particular, if we restrict $x$ and $x'$
to a section $\Sigma$ defined by $x=x_0$ and transform Eq.~(\ref{MCW}), 
we get,
\begin{eqnarray}\label{MCpartial}
   \left\langle \sum_n \Wt_n(x_0,x_0,y,p_y)\;\delta(E-E_n) \right\rangle
	& = &   \int \d p_x\; \delta(H(x_0,y,p_x,p_y)-E) \nonumber\\
        & = &   \sum  \left|\dydxv{H}{p_x}\right|^{-1} = 
\sum\frac{1}{|\dot{x}|},
\end{eqnarray}
where the sum is over all phase space points on the section $\Sigma$, 
$H=E$ with coordinates $(y,p_y)$. Note that this includes trajectories 
crossing $\Sigma$ in both senses.

This is not yet the result we need, because we have used the raw
wavefunctions instead of the renormalised versions in
Eq.~(\ref{decouplE}) to define the partial Wigner functions. The
transformations defined in Eq.~(\ref{defE}) are satisfactory for the
purposes of computing splittings because they conform intuitively to
the restriction and rescaling inherent in the Bogomolny transfer
operator \cite{bogomolnytrans} in the immediate neighbourhood of the
trajectory $\gamma_0$. They came naturally from Green's identity, but
we do not claim that they represent a clean, global definition of
Poincar\'e wavefunction (which should be made without reference to a
Green's function for example). Since we are not going to offer such a
definition, it is not worthwhile entering into a technically detailed
discussion of the effects of our particular rescaling, and instead we
will argue intuitively.

If we were to expand the wavefunctions in a basis of states with a
well-defined semiclassical interpretation (as WKB expressions using
Lagrangian manifolds for example), the Bogomolny rescaling is
semiclassically well-defined --- project out the components of the
basis states corresponding to crossing of $\Sigma$ in a particular
sense, and rescale the amplitudes by a square root of the transverse
velocity. The effect on the partial Wigner functions would be to
restrict them to oriented Poincar\'e sections and then to multiply by
the transverse velocity. As a result, we adjust Eq.~(\ref{MCpartial})
by restricting the sum to traversals of $\Sigma$ in a fixed sense and
multiplying each contribution by $v_x$. We assume that $(y,p_y)$
provide a single-valued parametrisation of the oriented Poincar\'e
section. Then, for each $(y,p_y)$, the microcanonical estimate is $1$
if $(y,p_y)$ is in the energetically allowed region of the plane and
is zero otherwise.  This is the content of Eq.~(\ref{MCPW}).

To summarise, we have argued that a coherent, global definition of
Poincar\'e-Wigner functions {\it should} have a microcanical
background given by Eq.~(\ref{MCPW}). We state without proof that our
particular definition, as implied by Eq.~(\ref{defE}), satisfies this
criterion near the intersection of $\gamma_0$ with $\Sigr_L$ (which is
the only place it is needed to calculate tunnelling). Global
properties, such as the nature of the transition from $1$ to $0$ at
the edge of the allowed region, are usefully defined only following a
more invariant definition, which is not offered.

\section{Inversion Symmetry}
\label{Invert}

As discussed in section~\ref{positivity}, positivity of 
splittings in a double well depends on the kind of 
symmetry underlying them. In two dimensions, for 
example, reflection-symmetric potentials $V(-x,y)=V(x,y)$
are responsible for positive splittings whereas inversion-symmetric 
potentials $V(-x,-y)=V(x,y)$ produce both positive and 
negative splittings. It turns out that we can see this dichotomy
very simply in the  example considered in Section \ref{thesystem},
which has both symmetries at once. It is instructive, as we do 
in this appendix, to investigate how the formalism depends
on which symmetry we use.

In Section \ref{thesystem} that problem was treated as if the symmetry
was reflection in $x$. As predicted in theory, all splittings were
then found to be positive. Had we instead defined $\R$ to be
inversion, however, we would have found negative splittings as a
result of the following simple observation.  A state that is called
``even'' with respect to reflection symmetry can be odd with respect
to inversion, and vice versa for an ``odd'' state. In such cases, a
splitting will change sign and become negative if we replace
reflection by inversion as the relevent symmetry.

To see in detail how this emerges, consider the table \ref{character}.
This shows the four classes of states, defined by how the transform
under the various symmetries.  For reflection, we define the
splittings to be the differences between the energies of class 2 and
those of class 1 or between the energies of class 4 and those of class
3, since in each case this is the odd energy minus the even energy
where odd/even is defined relative to the $\sigma_x$ operation. If
instead we define the splittings relative to $\sigma_x\sigma_y$, then
the table tells us that we should continue to use the differences
between classes 2 and 1 but should now take the differences between
the energies of class 3 and those of class 4. In other words, what we
were previously referring to as the odd $y$ parity doublets are now
responsible for negative splittings.

\begin{table}[ht]
\begin{center}
\begin{tabular}{|c|c|c|c|c|}
\hline
$$ & $I$ \hfill & $\sigma_x$ & $\sigma_y$ & $\sigma_x\sigma_y$\\
\hline
\hline
1 & + & + & + & +\\
2 & + & - & + & -\\
3 & + & + & - & -\\
4 & + & - & - & +\\
\hline
\end{tabular}
\end{center}
\vspace*{0.5cm}
\caption{\small
The four classes of state and how they transform under the
various symmetries.
}
\label{character}
\end{table}

In the calculation described in the text, this is evident if we look,
for example, at Eq.~(\ref{realHerring}), but having defined
$\psi_L(x,y)=\psi_R(-x,-y)$ rather than
$\psi_L(x,y)=\psi_R(-x,y)$. Since the states have definite $y$ parity,
this implies that there is simply an overall minus sign for the odd
parity states and no such sign for the even parity states.  How does
this sign manifest itself in the final results, such as
Eq.~(\ref{finally}) or Eq.~(\ref{absfinally})? The answer is
essentially that when we switch from $\sigma_x$ to $\sigma_x\sigma_y$
as the symmetry, we should change the sign of $W$, to account for a
reorientation of axes, and the positive-definite Gaussian
$g_M(\zetab)$ is replaced by a nondefinite one,
$g_{M'}(\zetab)$. Recall that the frame in which $W$ is calculated is
defined on $\Sigr_L$ to be the symmetric image of that on $\Sigl_R$,
and has opposite orientation according to whether we define
$\R=\sigma_x$ or $\R=\sigma_x\sigma_y$.

An added complication, which has already been alluded to, is that the
the matrix $W_0$ defining the function $f_0(E)$ similarly changes
sign. This makes perfect sense. Since some of the splittings are now
taken to be negative, we expect the average splitting value to be less
than before. Specifically, the fact that the coordinate $y$ has
changed sign implies that Eq.~(\ref{trunc}) becomes \cite{robbins}
\begin{eqnarray}
            f_0(E) 
& = & {1\over\pi} \frac{e^{-K_0/\hbar}} 
                      { \sqrt{\mbox{det} (W_0+I)}}\\
& = & {1\over\pi}  e^{-K_0/\hbar}
\left({\Lambda^{3/2} \over \Lambda^2-1} - 
     {\Lambda^{1/2} \over \Lambda^2-1}\right),
\end{eqnarray}
where $W_0$ is the monodromy matrix when the symmetry is $\sigma_x$
and $\Lambda$ its larger eigenvalue.  The second line follows from
expanding out the determinant assuming $\dim=2$ along the lines
discussed around Eq.~(\ref{stabfactor}). The two terms in brackets
have the interpretation of being the even and odd $y$ parity
contributions but now being subtracted rather than added, just as we
expect.

Note that this change in $f_0$ does not affect the magnitude of the
raw splittings obtained from Eq.~(\ref{enfin}). The modified value of
$f_0$ is compensated by the modified matrix $M$ so that the splittings
are the same with the exception that some of them are negative.
However, the normalised splittings are changed. This is clear. If they
should average to unity but some of them are now negative, then they
must all be rescaled to make this work out. This is implicit in
Eq.~(\ref{finally}) where the modified matrix $M$ entering the factor
$g_M$ is not compensated by the prefactor as in (\ref{enfin}).

Throughout this discussion, we have been comparing to the the picture
in which we used $x$ reflection to define the splittings. However, we
may not always have this choice. For example we could add a small term
to the potential such as $\epsilon xy$ which preserves the inversion
symmetry but breaks the reflection symmetry. If $\epsilon$ is small
enough then we do not expect the splittings to change very much.
Negative ones will mostly stay negative. So the possibility of
negative splittings is generic and not some pathology unique to this
problem. Another more interesting interaction which would maintain the
inversion symmetry while breaking the reflection symmetry would be to
add a uniform magnetic field. It is really that possibility which
motivates this appendix.

Also, this discussion has been specific to two dimensions. One could
imagine in higher dimensions that the symmetry involves reflecting
some of the coordinates on the section $\Sigma_L$ but not all of
them. Then the discussion at the beginning of the appendix still
applies if we change the sign of those variables which are reflected
and keep the sign of those which are not. It is too tedious to work
through all possibilities in a unified manner and we refrain from
doing so here.

\end{document}